\documentclass[journal, twocolumn, 10pt]{IEEEtran}

\usepackage{tabularx}
\newcolumntype{L}[1]{>{\raggedright\arraybackslash}p{#1}}
\newcolumntype{C}[1]{>{\centering\arraybackslash}p{#1}}
\newcolumntype{R}[1]{>{\raggedleft\arraybackslash}p{#1}}

\usepackage{color}
\usepackage{amsmath, amssymb, amsthm} 
\usepackage{url}
\usepackage{cite}	
\usepackage{mathtools}
\usepackage{footnote}
\usepackage{kbordermatrix}
\usepackage{dsfont}
\usepackage{enumitem}
\usepackage{dblfloatfix}
\usepackage{booktabs}

\ifCLASSINFOpdf
  \usepackage[pdftex]{graphicx}
\else
 
\fi
\ifCLASSOPTIONcompsoc
    \usepackage[caption=false, font=normalsize, labelfont=sf, textfont=sf]{subfig}
\else
\usepackage[caption=false, font=footnotesize]{subfig}
\fi
\linethickness{0.25mm} 
\makesavenoteenv{tabular}

\begin{document}
\title{Scalability Analysis of a LoRa Network under Imperfect Orthogonality}

\author{Aamir~Mahmood,~\IEEEmembership{Member,~IEEE},  
				Emiliano~Sisinni,~\IEEEmembership{Member,~IEEE}, Lakshmikanth~Guntupalli,~\IEEEmembership{Member,~IEEE},\\
        Ra{\'u}l Rond{\'o}n,
				Syed Ali Hassan,~\IEEEmembership{Senior Member,~IEEE}, 
        and~Mikael~Gidlund,~\IEEEmembership{Senior Member,~IEEE}% <-this % stops a space
\thanks{A. Mahmood, L. Guntupalli, R. Rond{\'o}n and M. Gidlund are with the Department of Information Systems and Technology, Mid Sweden University, 851 70 Sundsvall, Sweden, e-mail: aamir.mahmood@miun.se.}% <-this % stops a space
\thanks{E. Sisinni is with the Department of Information Engineering, University of Brescia, 25123 Brescia, Italy.}% <-this % stops a space
\thanks{S. A. Hassan is with the School of Electrical Engineering and Computer
Science, National University of Sciences and Technology, Islamabad 44000,
Pakistan.}% <-this % stops a space
\vspace{-15pt}
}



%\pagestyle{empty}
\maketitle

\begin{abstract}

Low-power wide-area network (LPWAN) technologies are gaining momentum for 
internet-of-things (IoT) applications since they promise wide coverage 
to a massive number of battery-operated devices using grant-free medium access. 
LoRaWAN, with its physical (PHY) layer design and regulatory efforts, has 
emerged as the widely adopted LPWAN solution. By using chirp spread spectrum 
modulation with qausi-orthogonal spreading factors (SFs), LoRa PHY offers 
coverage to wide-area applications while supporting high-density of devices. 
However, thus far its scalability performance has been inadequately modeled 
and the effect of interference resulting from the imperfect orthogonality of 
the SFs has not been considered. In this paper, we present an analytical 
model of a single-cell LoRa system that accounts for the impact of 
interference among transmissions over the same SF (co-SF) as well as 
different SFs (inter-SF). 
By modeling the interference field as Poisson point process under duty-cycled 
ALOHA, we derive the signal-to-interference ratio (SIR) distributions for 
several interference conditions. Results show that, for a duty cycle as low as 
0.33\%, the network performance under co-SF interference alone is considerably 
optimistic as the inclusion of inter-SF interference unveils a further drop 
in the success probability and the coverage probability of approximately 10\% 
and 15\%, respectively for 1500 devices in a LoRa channel. Finally, we illustrate how 
our analysis can characterize the critical device density with respect to 
cell size for a given reliability target.

\end{abstract}

%To this end, a scalability analysis is required to understand the performance 
%of LoRa in pragmatic conditions under interference.

%From the analysis, we calculate success probability, p_s, and coverage 
%probability, p_c. The obtaiend results demonstrate a significant impact of 
%inter-SF ineterfrence with a 10\% drop in p_s. Meanwhile, the results show 
%that the co-SF interference affects LoRa substantially by causing 15\% 
%decrement in p_c for 1500 devices per channel.


% For peer review papers, you can put extra informa tion on the cover
% \fi
\IEEEpeerreviewmaketitle

%%%%%%%%%%%%%%%%%%%%%%%%%%%%%%%%%%%%%%%%%%%%%%%%%%%%%%%%%%%%%%%%%%%%%%%%%%%
%%%%%%%%%%%%%%%%%%%%%%%%%%%%%%%%%%%%%%%%%%%%%%%%%%%%%%%%%%%%%%%%%%%%%%%%%%%
%%%%% Introduction
%%%%%%%%%%%%%%%%%%%%%%%%%%%%%%%%%%%%%%%%%%%%%%%%%%%%%%%%%%%%%%%%%%%%%%%%%%%
%%%%%%%%%%%%%%%%%%%%%%%%%%%%%%%%%%%%%%%%%%%%%%%%%%%%%%%%%%%%%%%%%%%%%%%%%%%
\section{Introduction} 

%\IEEEPARstart{O}{ver} the past few years, the implementation of internet- 
%of-things (IoT) in a variety of sectors, including smart- home, office, 
%city and industry, has experienced an exponential growth. In general terms, IoT 
%applications demand the network to be low cost, low energy, 
%reliable and, given the increasing number of connected devices, scalable. At 
%the moment, several technologies are available for IoT applications with 
%different limitations and levels of success. Zigbee, Bluetooth and WiFi are 
%some of the commonly used technologies for short-range communication, whilst 
%cellular networks (3G, LTE, 5G) are employed for long-reach scenarios.
\IEEEPARstart{O}{ver} the past few years, the implementation of internet- 
of-things (IoT) in a variety of sectors, including smart- home, office, 
city and industry, has experienced an exponential growth \cite{LRcommunicationStars}. 
In general, IoT applications require the network to be low-cost, low-energy, 
reliable and, given the increasing number of connected devices, scalable. At 
the moment, IoT applications are supported by available technologies such as 
%different limitations and levels of success. 
Zigbee, Bluetooth and WiFi %are some of the commonly used technologies
 for short-range communication; and cellular networks 
(e.g., 3G, LTE, 5G) for long-reach scenarios \cite{Hanke5G, zayas20183gpp}. 

However, applications that require the 
energy-efficient performance of short-range technologies but aim at reaching longer 
distances, prove to be a challenge for cellular systems. To this end, low-power wide-area networks (LPWAN) are recently taking the research 
spotlight \cite{UsmanLPWANReview}. LPWANs provide long-rage communication by limiting bit 
rates, making them a viable alternative for these cases. LoRaWAN is one of 
such emerging solutions supporting many smart applications \cite{UnderstandLoRaWANLimits}.

Smart metering, for instance, is one of the potential applications of 
LoRaWAN \cite{ICCSmartMetering}. The smart metering emables remote monitoring 
of resource (electricity, water, gas) consumption, which is becoming 
necessary both for operators and consumers, e.g., for demand 
response, dynamic pricing, load monitoring and forecasting 
\cite{Tripathi}\cite{SmartMeteringSurvey}. In urban cities, as the massive 
number of metering devices are to be connected, the choice of the 
communication 
technology depends on the number of supported devices and its scalability 
\cite{ASurveySmartGrid}. In this respect, LoRaWAN with its 
scalable--\textit{star-of-stars}--network architecture and simple medium access mechanism 
offers the required elements to support such applications. In a LoRaWAN cell, 
the nodes communicate with a gateway via a single-hop LoRa link using grant-free 
pure ALOHA protocol as medium access mechanism, which allows multiple nodes 
to uplink event-reports without any handshaking.
%%%%% before review 1
%In this respect, LoRaWAN with 
%its scalable network architecture and simple medium access mechanism offers 
%the required elements to support such applications. It operates in a star 
%topology where multiple nodes 
%communicate with a gateway using grant-free
%pure ALOHA protocol as medium access mechanism, which allows the devices to uplink event-reports without 
%any handshaking. 
%%%%%%%
As ALOHA works without any contention mechanism 
and is prone to collisions, LoRa--the physical (PHY) layer 
\cite{semtech120022}--offers degrees of freedom in carrier 
frequency (CF), bandwidth (BW), coding rate (CR) and spreading factors (SFs) 
to orthogonalize transmissions, and thus can support high-density deployment 
of devices. Also, thanks to the chirp spread spectrum (CSS) modulation employed by LoRa, the SFs act as virtual 
channels for a given CF, BW and CR. Despite the apparent robustness of LoRaWAN technology, a 
scalability analysis; \textit{how does LoRa scale with device density and cell size while ensuring a 
certain quality of service}, is required to understand its potential for enabling 
smart applications. 

\subsection{Related Works and Motivation}
In general, the scalability of a LoRa network can be affected by: 
a) \textit{co-SF interference}, caused by unruly same channel transmissions 
using the 
same SF, that can restrict the scalability in plausible high-density 
deployments if the signal-to-interference ratio (SIR) of the desired 
transmission is below a certain threshold, b) \textit{inter-SF interference}, 
which stems from the imperfect orthogonality among different SFs and implies 
that the transmissions from different SFs are not completely immune to 
the adjacent SFs, thus requiring a certain level of SIR protection. 
%1) \textit{co-SF interference}, which is caused by 
%unruly same channel transmissions using the same SF. Unless the 
%signal-to-interference ratio (SIR) of the desired transmission is above a certain 
%threshold, the concurrent transmissions in plausible high-density deployments 
%can restrict the scalability
A common assumption in the literature is that the different SFs are completely 
orthogonal to each other, thus providing complete protection to concurrent 
transmissions of different SFs \cite{georgiou2017low}. Recently in 
\cite{ImpfectOrthogo}, based on simulations and USRP-based implementation, 
the impact of quasi-orthogonality of the SFs on link-level performance 
reveals the opposite, thus invalidating that common notion. Although the required SIR 
protection is low, it can severely affect the network performance  
in cases when the interfering terminal is closer to the receiver (a 
gateway) than the desired terminal, usually referred to as the \textit{near-far} 
conditions.

Apart from assuming the complete orthogonality among SFs, the LoRa 
performance modeling and analysis under co-SF interference is also limited.  
In \cite{BorVoigt, LoRaSensors, NS3Simulations, ICCSmartMetering}, the 
authors analyzed the LoRa  performance only by simulations. 
On the other hand, there exists a few analytical studies in the literature, for 
instance, in \cite{LoRaSensors, Mikhaylov} 
%Whereas, in a few 
%existing analytical studies e.g., 
%\cite{LoRaSensors, Mikhaylov, UnderstandLoRaWANLimits, georgiou2017low}, the 
%impact of interference is modeled inadequately. In 
%\cite{LoRaSensors, Mikhaylov}, 
pure ALOHA was assumed and the concurrent 
transmissions were considered to be lost regardless of the SIR level at the 
receiver. Similarly in \cite{UnderstandLoRaWANLimits}, LoRa capacity was 
studied based on the superposition of independent pure ALOHA-based virtual 
networks corresponding to the available SFs per channel. In 
\cite{georgiou2017low}\cite{ModelCaptureEffect}, on the contrary, the 
performance of a LoRa 
system was analyzed under the capture effect, i.e., the SIR of a desired 
signal is above an isolation threshold for successful packet reception. 
However in \cite{georgiou2017low}, only the strongest co-SF interferer was 
modeled using stochastic 
geometry. While in \cite{ModelCaptureEffect}, the model considered the 
capture effect with each co-SF interferer separately wherein the channel fading was also ignored. Therefore, the system level performance of a LoRa network under the aggregate effect of both the 
co-SF interference and 
inter-SF interference is yet to be modeled and investigated.

\subsection{Contributions}

In this paper, we investigate the scalability of a LoRa network--a 
multi-annuli single cell system--based on uplink coverage probability under the 
joint effect of both the co-SF and inter-SF interference. The multi-annuli 
structure is a simple yet logical scheme for allocating SFs based 
on their respective signal-to-noise ratio (SNR) thresholds. In addition, it 
helps to realize the near-far conditions and hence to analyze the impact of 
inter-SF interference. We apply the tools from stochastic geometry to model the 
interference field as a Poisson point process (PPP), and include the medium 
access and control (MAC)- and PHY-layer features and regulatory constraints 
into the model. Thereon, the key contributions of this paper are summarized as follows:
%We derive the SIR 
%distributions under a realistic path loss model and fading in the presence of 
%co-SF interference for two cases, a) dominant interferer and b) aggregate 
%interference power, and also under inter-SF interference. Using these 
%results, we first show how two SIR distributions for co-SF interference 
%differ with respect to device density. In general, our analysis reveals 
%that although the co-SF interference plays dominant role in coverage loss in 
%interference-limited scenarios, inter-SF interference also has a significant 
%impact. It 
%by affecting the outage probability of especially for higher SFs, further 
%reduces the coverage probability by 15\% for a small number (considering the 
%extreme duty-cycle limitations) of end-devices. We also study the network 
%scalability, under various medium access 
%restrictions, using coverage probability contours that help to characterize 
%the critical node density and cell size under reliability constraints. 
%Furthermore, the analytical results are validated using Monte Carlo 
%simulations and both match precisely indicating the accuracy of the analysis.
%%
%%\textcolor{red}{The key contributions of this paper are summarized as follows:
\begin{itemize}%[leftmargin=*]
	\item The SIR distributions are derived in the presence of dominant co-SF 
interferer, cumulative co-SF interference and inter-SF interference, under a 
realistic path loss model and channel fading. It is shown how two co-SF 
interference-based 
distributions differ; the dominant interferer 
case, giving an upper bound on success probability, loses its imperviousness 
to cumulative co-SF interference effects with an increase in device density.  
	\item Using SIR distributions, the coverage probability is evaluated. It is 
shown that co-SF interference causes a major coverage loss in 
interference-limited scenarios. However due to imperfect orthogonality, 
inter-SF 
interference exposes the network for further 15\% coverage loss for a small 
number of concurrently transmitting end-devices.
	\item Coverage probability contours are presented, which accentuate the 
significance of our analytical models for scalability analysis and also can act as 
a valuable tool for dimensioning the cell size and node 
density under medium access and reliability constrains.
	\item How a strategy to allocate SFs to the devices influences the overall network 
performance is studied using three different SF allocation 
schemes.
	\item A extension framework for modeling interference in a multi-cell 
network is formulated based on the presented single-cell model.
%\item The analytical results are validated using Monte Carlo simulations and both 
%match precisely indicating the accuracy of the analysis.
\end{itemize}
%We also introduce coverage probability contours for possible LoRa channel 
%settings, which can give a simple scalability measure under outage constraints.

%- If the overlap is not strong enough, the packets are not necessarily lost due 
%to capture effect.

The rest of this paper is organized as follows. Section~\ref{sec:LoRa_and_WAN} 
introduces the LoRa system. Section~\ref{sec:SystemModel} presents the 
network geometry and, signal and channel models. 
Section~\ref{sec:OutageNoInterference} finds the SNR-based success 
probability while Section~\ref{sec:OutageInterference} develops the SIR 
distributions under interference. Section~\ref{sec:SimulationResults} presents numerical results, analyzes the effect of 
different SF-allocation schemes and gives a framework to extend our model to a 
multi-cell LoRa network. Finally, 
Section~\ref{sec:Conclusions} concludes the paper.

%%%%%%%%%%%%%%%%%%%%%%%%%%%%%%%%%%%%%%%%%%%%%%%%%%%%%%%%%%%%%%%%%%%%%%%%%%%
%%%%%%%%%%%%%%%%%%%%%%%%%%%%%%%%%%%%%%%%%%%%%%%%%%%%%%%%%%%%%%%%%%%%%%%%%%%
%%%%%%%%%%%%%%%%%%%%%%%%%%%%%%%%%%%%%%%%%%%%%%%%%%%%%%%%%%%%%%%%%%%%%%%%%%%
%%%%%%%%%%%%%%%%%%%%%%%%%%%%%%%%%%%%%%%%%%%%%%%%%%%%%%%%%%%%%%%%%%%%%%%%%%%
\section{The LoRa System}
\label{sec:LoRa_and_WAN}
%%%%%%%%%%%%%%%%%%%%%%%%%%%%%%%%%%%%%%%%%%%%%%%%%%%%%%%%%%%%%%%%%%%%%%%%%%%
%%%%%%%%%%%%%%%%%%%%%%%%%%%%%%%%%%%%%%%%%%%%%%%%%%%%%%%%%%%%%%%%%%%%%%%%%%%
%%%%%%%%%%%%%%%%%%%%%%%%%%%%%%%%%%%%%%%%%%%%%%%%%%%%%%%%%%%%%%%%%%%%%%%%%%%
%%%%%%%%%%%%%%%%%%%%%%%%%%%%%%%%%%%%%%%%%%%%%%%%%%%%%%%%%%%%%%%%%%%%%%%%%%%

\begin{figure}[!t]
	\centering
		\includegraphics[width=0.95\linewidth]{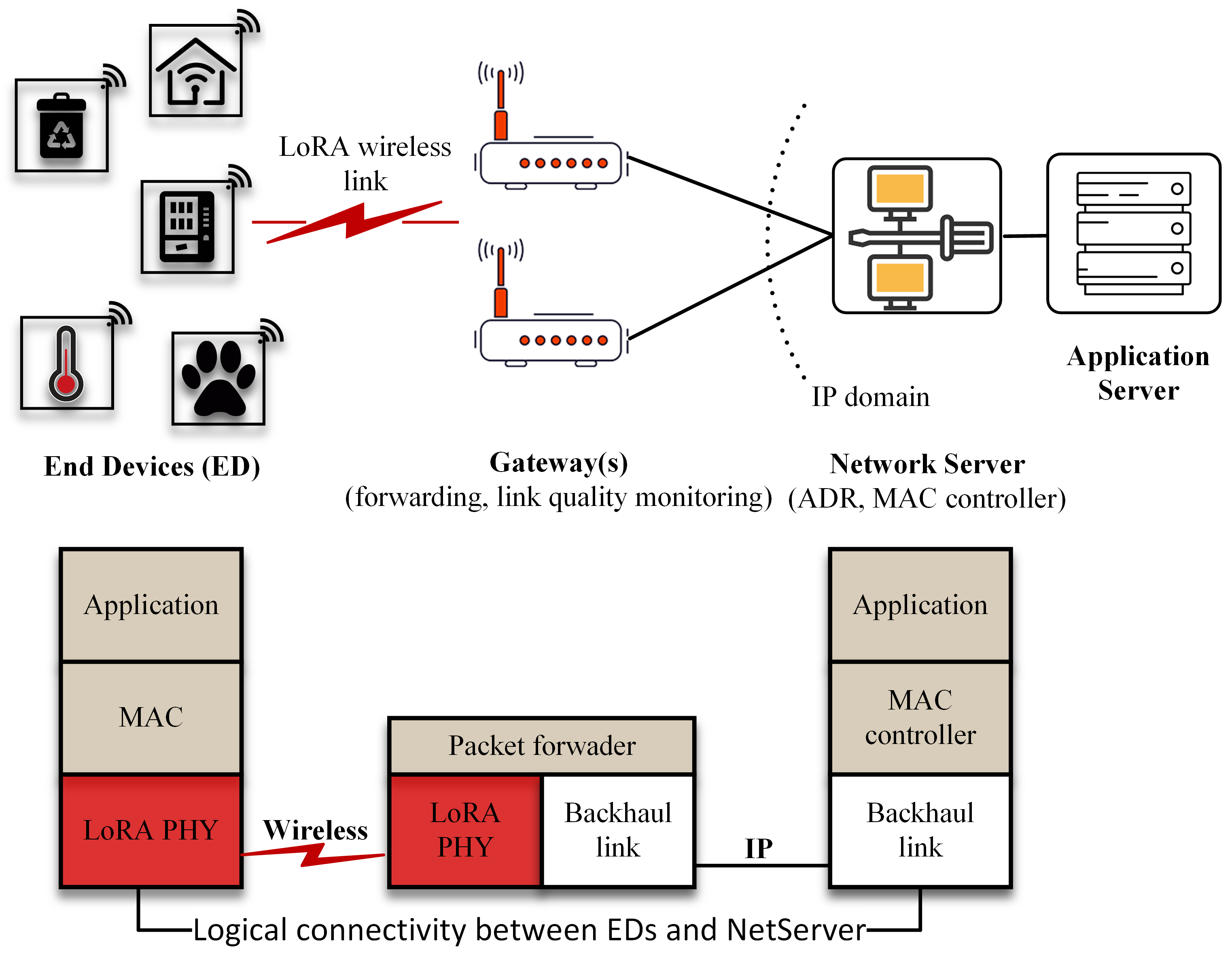}
		\vspace{-8pt}
	\caption{Overview of the LoRaWAN architecture}
	\vspace{-12pt}
	\label{fig:LoRaWAN_arch}
\end{figure}

As an LPWAN solution, the LoRa system consists of two main components: LoRa--a 
proprietary PHY layer modulation scheme designed by 
Semtech Corporation \cite{LoRaPatent}, and LoRaWAN--the rest of the protocol stack 
developed, as an open standard, by LoRa Alliance 
\cite{LoRaWANSpecs}. Fig.~\ref{fig:LoRaWAN_arch} summarizes the LoRaWAN architecture and the 
protocol stack.     

%The LoRaWAN is an open standard promoted by the LoRa alliance \cite{LoRaWANS pecs}; it belongs to 
%the low power wide area network (LPWAN) family, i.e. those wireless 
%communication solutions aiming to mimic cellular network arrangement still 
%offering a reduce power by means of a relatively small bit rate. The cellular 
%architecture (with a base station covering a large geographical area) well 
%fits typical internet-of-things (IoT) requirements. Formally, the LoRaWAN 
%specs only covers the data link layer. The underlying physical layer 
%(PHY) exploits proprietary transceivers developed by Semtech. The star topology 
%inherited by cellular approach avoids a true network layer and the 
%application layer is user specific. The network follows a hierarchical 
%organization, which allows innovative business models, since the owners of the 
%nodes, infrastructure and the transferred information can all be different.

%unlicensed sub-GHz frequency bands

\subsection{LoRaWAN Architecture}
\label{subsec:LoRaWAN}

A LoRa network provides wireless connectivity analogous to cellular 
systems but optimized in terms of energy efficiency for IoT-focused 
applications. The network typically follows a hierarchical or 
\textit{star-of-stars} architecture, where the end devices (EDs) are 
connected via LoRa PHY to one or many gateways. The gateway
%, also called as 
%base station or packet forwarder, 
is then connected to a common 
network server (NetServer) over standard IP protocol stack. The NetServer 
is finally connected to an application server (AppServer) over IP. The functionality 
of each entity is as follows: 
\begin{itemize}[leftmargin=*]
	\item \textit{End device (ED)} supports both the uplink and downlink messages to/from the 
gateway, generally with a focus on event-triggered uplink transmissions.

	\item \textit{Gateway} performs relaying of the messages to EDs or NetServer 
received over LoRa PHY or IP interface. The gateway is 
transparent to the EDs, which are logically connected to the NetServer.

	\item \textit{NetServer} handles the overall network management, e.g.,  
resource allocation (such as SF and bandwidth) for enabling 
adaptive data rate (ADR), authentication of EDs etc.  

	\item \textit{AppServer} is in charge of admitting EDs to the network and taking care of 
data encryption/decryption.
\end{itemize}

The LoRa network operates on unlicensed sub-GHz RF bands, which are subject 
to regulation on medium access duty-cycling or listen-before-talk (LBT) and 
effective radiated power (ERP). The most common 
approach for the wireless medium access is the simple ALOHA 
protocol, which is primarily regulated by the NetServer.

\subsection{LoRa PHY layer}
\label{subsec:LoRaPhysical}

The LoRa PHY--a derivative of CSS modulation--spreads the data symbols with chirps, where each chirp is  
a linear frequency-modulated sinusoidal pulse of fixed bandwidth $B\! =\! f_1\! -\! f_
0$ and chirp duration $T_c$. By varying the chirp duration, 
quasi-orthogonal signals, acting as virtual channel, can be 
created.
 %to overcome the spectrum 
%scarcity in unlicensed bands. 
In addition, the chirp duration leads to a 
tradeoff between the throughput and the robustness against noise and 
interference. For a fixed $T_c$, the data symbols are coded by 
unique instantaneous frequency trajectory, obtained by cyclically shifting a 
reference chirp. These cyclic shifts, representing symbols, are discretized 
into multiples of chip-time $T_{\mathrm{chip}} = 1/B$, while only $2^{j}$ 
possible edges in the instantaneous frequency exist. 
Therefore, each chirp represents $j$ bits where $j$ is referred to as 
SF. As a result, the modulation signal, $m(t)$, of $n$th 
LoRa symbol can be expressed as
\vspace{-4pt}
\begin{equation}
  m(t) = 
    \begin{cases}
      f_1 + k \cdot \left(t - nT_{\mathrm{chip}}\right) & \text{if $0 \leq t \leq nT
_{\mathrm{chip}}$}\\
      f_0 + k \cdot \left(t - nT_{\mathrm{chip}}\right) & \text{if $nT_{\mathrm{chip
}} < t \leq T$}, \nonumber
    \end{cases}       
\end{equation}
where $k=(f_1 - f_0)/T_c$ is the rate of frequency increase over symbol 
duration $T_c$.

The NetServer can adapt the data rate
 %according to the transmission 
%conditions 
by changing the bandwidth $B \in \{
125, 250\}$ kHz and $\textrm{SF} \in \{7,\cdots, 12\}$, which together relate to chirp duration as $T_c = 2^{\textrm{SF}}/B$. Note that the chirp 
rate remains the same, and equals to $B$, while the chirp duration (consequently time-on-air) increases drastically with the SF. 
%Considering the 
%possible coding rate ($\textrm{CR} = N/M$, with $M 
%= \{5,\cdots,8\}$ the codeword length and $N=4$ the data block length) the 
%raw over the air bit rate 
%$R_b = \textrm{SF}\cdot \frac{B}{2^\mathrm{SF }}\cdot\textrm{CR}$ 
%can vary from about 180 bps ($B = 125$ kHz, $\textrm{CR} =  4/8$ and $\textrm{
%SF} = 12$) to about 11 kbps ($B = 250$ kHz, $\textrm{CR}=4/ 5$ and $\textrm{SF
%} = 7$). 
On the positive side, higher SF yields higher processing gain and 
thus reduces the target signal-to-noise ratio (SNR) for correct reception at the 
receiver. 
%In other words, higher SFs can be utilized to extend the 
%communication range of the EDs. 
%\begin{figure}[!t]
	%\centering
		%\includegraphics[width=0.65\linewidth]{SystemModel.png}
	%\caption{Single gateway LoRa network model with concurrently 
%active nodes in the regions employing same or different SFs.}
	%\label{fig:systemModel}
%\end{figure}
%\begin{figure}[!t]
	%\centering
		%\includegraphics[width=0.9\linewidth]{LoRaMAP1.png}
	%\caption{Allocation of SFs under realistic signal propagation 
%conditions. This figure is a modified version of \cite{atdi_LoRaMAP} to 
%develop a theoretical 
%model close to the practical deployment of a LoRa network.}
%\vspace{-15pt}
	%\label{fig:systemModelATDI}
%\end{figure}

Because of ALOHA-based MAC, two or more signals from EDs using the same or 
different SFs can overlap in time and frequency. In such cases, the demodulator output can be indistinct depending on the 
isolation threshold and the processing gain of the SFs. The 
signals using $\textrm{SF}=i$ and $\textrm{SF}=j$ can be decoded correctly only if capture effect \cite{MultiHopLoRa} occurs, i.e., the SIR of the desired signal is above the isolation threshold. The thresholds for 
two conditions, co-SF interference $i = j$ and inter-SF interference $i \neq j
$, are given by the SIR matrix \cite{ImpfectOrthogo}

\vspace{-12pt}
\renewcommand{\kbldelim}{[}% Left delimiter
\renewcommand{\kbrdelim}{]}% Right delimiter
\begingroup
\small
\renewcommand*{\arraystretch}{1.25}%
\begin{align}
  \mathbf{\Delta}_{[dB]} \!= \!\!\kbordermatrix{
    & \textrm{SF}_7 & \textrm{SF}_8 & \textrm{SF}_9 & \textrm{SF}_{10} & \textrm{SF}_{11} & \textrm{SF}_{12}\cr
              \textrm{SF}_7 \!\!\!     & \!1   & -8  & -9  & -9  & -9  & -9  \cr
              \textrm{SF}_8 \!\!\!     & \!-11  & 1   & -11  & -12  & -13  & -13  \cr
							\textrm{SF}_9 \!\!\!     & \!-15 & -13 & 1   & -13 & -14 & -15 \cr
							~\textrm{SF}_{10} \!\!\! & \!-19 & -18 & -17 & 1   & -17 & -18 \cr
							~\textrm{SF}_{11} \!\!\! & \!-22 & -22 & -21 & -20 & 1   & -20 \cr
							~\textrm{SF}_{12} \!\!\! & \!-25 & -25 & -25 & -24 & -23 & 1 \cr}\,\,\,\,\,\,
\label{eq:sinrMatrix}
\end{align}
\normalsize
\endgroup
Each element $\delta_{ij}[\textrm{dB}]$ in $\mathbf{\Delta}$ is the SIR margin that a 
packet sent at $\textrm{SF}_i$ must have in order to be decoded correctly if the 
colliding packet is sent at $\textrm{SF}_j$.

\begin{figure}[!t] 
    \centering
  \subfloat[]{%
       \includegraphics[width=0.43\linewidth]{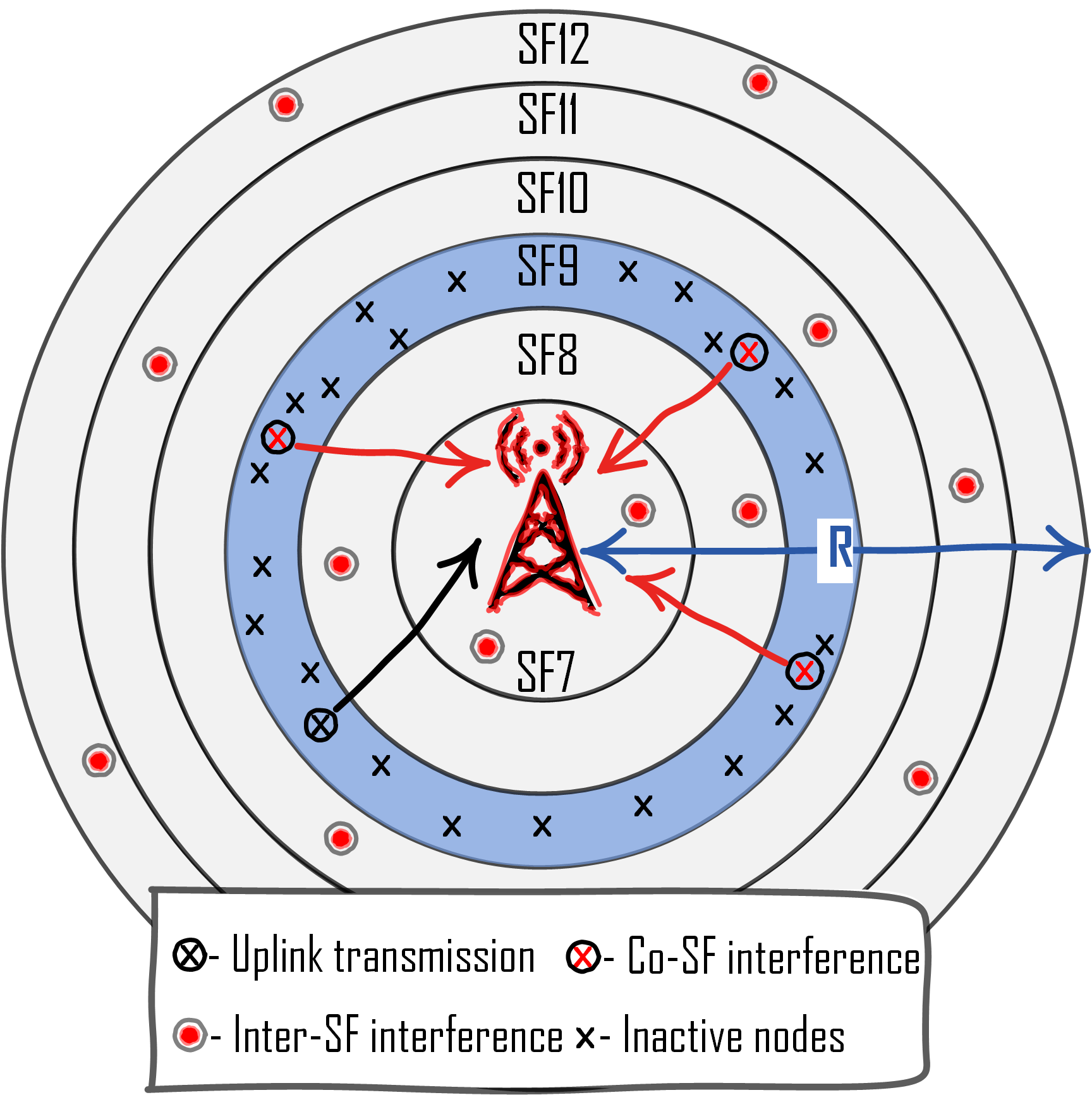}\label{fig:1a}}\hspace{0.5em}%\hfill
  \subfloat[]{%
        \includegraphics[width=0.55\linewidth]{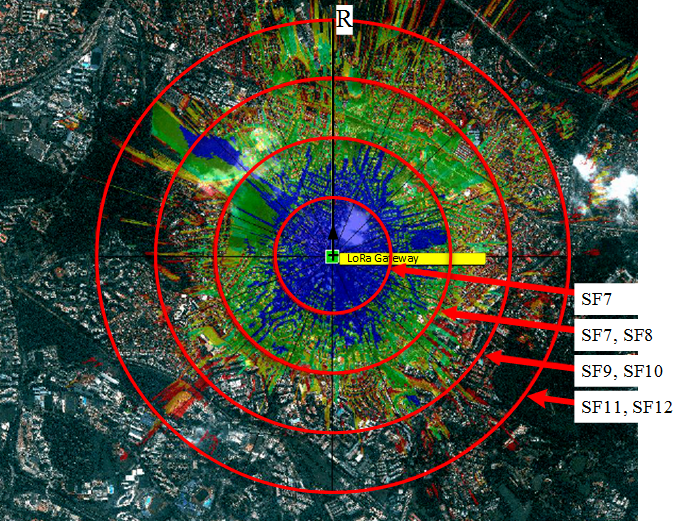}\label{fig:1b}}
	\vspace{-4pt}
  \caption{System model: (a) single gateway LoRa network model with concurrently 
active nodes in the regions employing same or different SFs, (b) allocation of SFs under realistic signal propagation 
conditions. This figure is a modified version of \cite{atdi_LoRaMAP} to 
develop a theoretical 
model close to the practical deployment of a LoRa network.}
\vspace{-10pt}
  \label{fig:systemModel}      
\end{figure}

%%%%%%%%%%%%%%%%%%%%%%%%%%%%%%%%%%%%%%%%%%%%%%%%%%%%%%%%%%%%%%%%%%%%%%%%%%%
%%%%%%%%%%%%%%%%%%%%%%%%%%%%%%%%%%%%%%%%%%%%%%%%%%%%%%%%%%%%%%%%%%%%%%%%%%%
%%%%%%%%%%%%%%%%%%%%%%%%%%%%%%%%%%%%%%%%%%%%%%%%%%%%%%%%%%%%%%%%%%%%%%%%%%%
%%%%%%%%%%%%%%%%%%%%%%%%%%%%%%%%%%%%%%%%%%%%%%%%%%%%%%%%%%%%%%%%%%%%%%%%%%%
\section{System, Signal and Channel Models}
\label{sec:SystemModel}
%%%%%%%%%%%%%%%%%%%%%%%%%%%%%%%%%%%%%%%%%%%%%%%%%%%%%%%%%%%%%%%%%%%%%%%%%%%
%%%%%%%%%%%%%%%%%%%%%%%%%%%%%%%%%%%%%%%%%%%%%%%%%%%%%%%%%%%%%%%%%%%%%%%%%%%
%%%%%%%%%%%%%%%%%%%%%%%%%%%%%%%%%%%%%%%%%%%%%%%%%%%%%%%%%%%%%%%%%%%%%%%%%%%
%%%%%%%%%%%%%%%%%%%%%%%%%%%%%%%%%%%%%%%%%%%%%%%%%%%%%%%%%%%%%%%%%%%%%%%%%%%
\subsection{System Model}
\label{subsec:SysGeom}
We present the uplink system model for a single LoRa gateway, which takes into 
account the interference from concurrent uplink transmissions on a desired 
uplink transmission. The notation used in this paper are 
summarized in Table~\ref{tab:symbols}.
The system model is as follows:
\begin{itemize}[leftmargin=*]
\item End devices are spatially distributed in a deployment region $\mathcal
{S} \subseteq \mathbb{R}^2$, which is a 2-D Euclidean space, according to a 
homogeneous PPP $\Phi$ of intensity (density) $\lambda>0$, with a gateway 
located at its origin. The 
region $\mathcal{S}$ is a disk of radius $R$ and area $A =|\mathcal{S}| = \pi 
R^2$, and it contains $\Phi(A)$ devices which is a Poisson random variable 
with mean $\bar{N} = \lambda A$.

\item Devices make independent decisions to transmit in the uplink using 
ALOHA. In addition, the devices satisfy the per-frequency band duty cycle
constraint of $\alpha$, as per ETSI specifications \cite{LoRaWANSpecs}. Therefore, the 
set of all transmitting devices at a given time also makes a homogeneous PPP $\Phi_m$ of 
intensity $
\alpha\lambda$ due to independent thinning of the PPP.

\item Each device is equipped with an omni-directional antenna, and transmits 
at a fixed transmission power $p_t$ in a same channel of bandwidth $B$. 

\item Each device is assigned an SF by the NetServer 
according to its distance from the gateway. In reality, it can be assigned 
based on the SNR of the received packets, and the devices in a certain 
range can use the same SF (see Section \ref{subsec:SFAStrategy}). For simplicity, the network is divided 
into $K$ disjoint annuli of width $r_i = R/K$ each, starting at 
the center and moving outwards. In this case, if $\mathcal{S}_i$ are disjoint subsets of $\mathcal{S}$, then $\Phi_m$ is the superposition 
of $\Phi_{m,i}~(i=1, \cdots, K)$. The parameter $K$ is determined by the 
cardinality 
of the set of available SFs, that is, $K=\lvert \textrm{SF} \rvert
$. The annuli set is denoted as $\mathcal{K} = \{1,\cdots, K\}$, and $i$th 
($1\leq i \leq K$) annulus defined by the inner and outer radii $\ell_{i-1}$ 
and $\ell_{i}$ have the same SF, and the area of the $i$th 
annulus is $a_i = \pi r_i^2 \left(2i-1\right)$.
%For example, according to these definitions, the device located within 
%annulus 1 with boundaries $\ell_0$ and $\ell_1$ use ${\textrm{SF}_7}$ and so 
%on. 
%Note that the recommended value of $r_c$ by Semtech is $2$ km \cite{LoRaRange}.
\end{itemize}

An illustration of a multi-annuli single gateway LoRa network with 
concurrently active nodes in the regions employing same SF as 
well as different SFs is shown in Fig.~\ref{fig:systemModel}(a). 
The simultaneous same SF transmissions interfere at the gateway to a desired transmission thus resulting in 
co-SF interference, whereas the inter-SF interference 
results from concurrent transmissions from all the other regions with devices 
employing quasi-orthogonal SFs. Note that in realistic signal propagation 
conditions, the sharp boundaries for SFs allocation as in 
Fig.~\ref{fig:systemModel}(a) might not exist. In practice, it is likely to have 
mixed SNR levels favoring the allocation of two SFs in the same annulus as 
in Fig.~\ref{fig:systemModel}(b). However, our mathematical model can easily 
address any of such network configurations without loss of generality.    

\begin{table}[!t]
\centering
                \footnotesize
                \caption{List of Important Notations}
                \begin{center}
                \hspace*{0.3cm}\begin{tabular}{R{1.0cm}||p{6.5cm}}
                    %\hline
										\noalign{\hrule height 1pt}
                    \rule{0pt}{10pt}\textbf{Symbol}  & \textbf{Description} \\
										\noalign{\hrule height 1pt}
                    $\mathbb{R}^2$   & 2-dimensional Euclidean space \\
										$\alpha$         & Duty cycle constraint \\
                    $\Phi, \Phi_m$   & PPP of EDs, active EDs with intensity $\lambda, \alpha\lambda$\\ 
										%$$			         & PPP of concurrently active EDs with intensity $\alpha\lambda$\\ 
										$K, \mathcal{K}$ & Total number of annuli, set of annuli $\{1,\cdots, K\}$\\ 
                    %$\mathcal{K}$    & Set of annuli $\{1,\cdots, K\}$ \\
										$a_i, r_i$       & Area, width of $i$th annulus \\
										$\ell_i, \ell_{i-1}$ & Outer, inner boundary of $i$th annulus \\
										$v_i$            & Average number of active EDs in $i$th annulus\\
										$\eta$           & Path loss exponent \\
										$\theta_{\textrm{SF}}$  & SNR threshold for an $\textrm{SF}$ \\
                    $\delta_{ij}$    & SIR threshold for successful reception of a packet transmitted at $\textrm{SF}_i$ under a concurrent transmission at $\textrm{SF}_j$ \\
                    $h(t), g(t)$     & Fading coefficient of the useful, interfering signal \\
										$H, G$           & Channel gain of the useful, interfering signal \\
										$\mathbb{E}[\cdot], \mathbb{P}[\cdot]$ & Statistical expectation, probability measure\\
                    $\sigma^2$       & Variance of AWGN noise \\
                    $N_{\textrm{SF}_x}$ & Total number of EDs using $\textrm{SF}_x$\\
                    $l(x_i)$         & Free space path loss of an $i$th ED located at $x$ \\             
                    $\mathds{1}_i^{\textrm{SF}_x}$   & Indicator function of $i$th ED using $\textrm{SF}_x$\\%, which takes the value 1 when the node is active and 0 otherwise\\
										%$F_X(x)$         & CDF of a random variable $X$ \\
                    $P_{\textrm{SNR}}$      & Success probability in AWGN noise \\
										$P_{\textrm{SIR}}^*$    & Success probability under dominant co-SF interferer\\
                    $P_{\textrm{SIR}}$      & Success probability under cumulative co-SF interference\\
										$P_{\textrm{SIR}}^{\Pi}$ & Success probability under cumulative co-SF \& inter-SF interference\\
                    $\mathcal{I}_{\textrm{CSF}}$ & Total interference from concurrently active co-SF EDs  \\
										$\mathcal{I}_{\textrm{ISF}}$ & Total interference from concurrently active inter-SF EDs\\
                    $\mathcal{L}_{\mathcal{I}_{\textrm{CSF}}}(\cdot)$  & Laplace transform of $\mathcal{I}_{\textrm{CSF}}$  \\ 
																$\mathcal{L}_{\mathcal{I}_{\textrm{ISF}}}(\cdot)$  & Laplace transform of $\mathcal{I}_{\textrm{ISF}}$  \\
                    $P_c[\mathcal{Y}]$   & Coverage probability with $\mathcal{Y}\!\in\! \{P_{\textrm{SNR}}, P_{\textrm{SIR}}^*, P_{\textrm{SIR}}, P_{\textrm{SIR}}^{\Pi}\}$ \\
                \noalign{\hrule height 1pt}
                \end{tabular}
								\end{center}
								\vspace{-20pt}
								\label{tab:symbols}
\end{table}

\subsection{Signal and Channel Model}
\label{subsec:SignChanModel}
While focusing on a single end device, we investigate its uplink 
performance under the simultaneous interfering transmissions originating from 
co-SF and inter-SF regions. Although the inter-SF transmissions are less disruptive (i.e., requiring less SIR 
threshold), the impact of inter-SF interference must be 
taken into account for a realistic analysis. Assume that device $x_1$ located in $i$th 
annulus intends to communicate with the gateway, and the devices other than 
$x_1$ act as a set of potential interferers. Let $s_1\left(
t, \textrm{SF}_p\right)$ be the desired signal transmitted with spreading factor $ \textrm{SF}_p$ and experiences block-fading  with instantaneous fading 
coefficient $h(t)$ in addition to power-law path loss. The $h(t)$ is a zero-mean circularly symmetric complex 
Gaussian (CSCG) random variable with unit-variance 
i.e., $\mathbb{E}[\lvert h \rvert^2] = 1$ which corresponds to Rayleigh 
fading. Similarly, let $s_i\left(t, \textrm{SF}_q\right)$ be the interfering 
signal from device $i$ at $\textrm{SF}_q$ over the Rayleigh 
block-fading channel with fading coefficient $g(t)$. Then the received 
signal, $r_1(t)$, under co-SF and inter-SF
interference can be expressed as
\vspace{-8pt}
\begin{align} 
r_{1}(t)\! & = \! l(x_{1}) h(t) \! \ast \! s_{1}(t, \textrm{SF}_p) \! + \!\!\! \sum_{j=2}^{N_{\textrm{SF}_{p}}} \!\mathds{1}_j^{\textrm{SF}_p} l(x_j) g_j(t) \! \ast \! s_j(\!t, \textrm{SF}_p\!) \nonumber \\[-7pt]
				& + \!\!\!\!\sum_{q\in \mathcal{K} \setminus p}\bigg({\sum_{k=1}^{N_{\textrm{SF}_{q}}} \mathds{1}_k^{\textrm{SF}_q} l(x_k) g_k(t) \ast s_k(t, \textrm{SF}_q)}\bigg) + n(t).
\label{eq:signalModel}
\end{align}

In this signal model, other than explaining the notations a few remarks are:
\begin{itemize}[leftmargin=*] 	

\item $n(t)$ is the additive white Gaussian noise (AWGN) 
with zero mean and variance $\sigma^2 = N_0 + \textrm{NF} + 10\log_{10} B \,[
\textrm{dBm}]$, where $N_0$ is the noise power density, NF is the receiver 
design-dependent noise figure, and $B$ is the channel bandwidth. 
%Note that NF is the 
%radio-design dependent and it is fixed for a given implementation.  

\item $\mathds{1}_i^{\textrm{SF}_x}$ is an indicator function for 
a device $i$ transmitting at $\textrm{SF}_x$, and $N_{\textrm
{SF}_{x}}$ is the total number of devices using $\textrm{SF}_{x}$.

\item $l(x_i)$ is the path loss attenuation function, where $x_i$ is the 
Euclidean distance in meters between the device $i$ and the gateway. From the Friis 
transmission equation and following a non-singular model \cite{aljuaid2010}, 
we consider $l(x_i) = \kappa \left[\max\left(x_i, x_c\right)
\right]^{-\eta}$, where $\eta$ is the path loss exponent, and $x_c > 0$ is the critical distance to avoid that $l(x_i)$ 
tends to infinity (i.e., when $x_i \rightarrow 0$). In addition, $\kappa = (
\lambda_c/4\pi)^2$ where $\lambda_c$ is the carrier wavelength.     
\end{itemize}

\subsection{Performance Metrics}

%In the signal model \eqref{eq:signalModel}, the second term counts the co-SF 
%interferers while the third considers the concurrent inter-SF transmissions. 
In absence of any interference, the link performance is determined by 
an SF specific SNR threshold. On the other hand, when co-SF and/or inter-SF 
transmissions interfere with the desired signal, the performance can be 
determined by the SF-dependent SIR. In this respect, the cumulative 
distribution function (CDF) of SNR/SIR is an important measure for 
characterizing the link performance; that is, for a given SNR/SIR threshold $
\tau$, the CDF gives the link outage probability $P_o$, whereas the 
complementary CDF (CCDF) gives the success probability 
$P_{\mathcal{X}}$, $P_{\mathcal{X}} = 1 - P_o$. If $\mathcal{X} \in \{\mathrm{SNR}, \mathrm{SIR}\}$, then 
\begin{equation}
P_{\mathcal{X}} = \mathbb{P}\left[\mathcal{X} \geq \tau \right].
\label{eq:PM1}
\end{equation}

The other important metric, derived from the $P_{\mathcal{X}}$, is the coverage 
probability $P_c$, which is equivalent to the probability that a randomly chosen 
device achieves the target SNR/SIR threshold $\tau$. It can be defined as
\begin{equation}
P_c = \mathbb{E}_D\Big[\mathbb{P}\left[\mathcal{X} > 
\tau | D = x_{1} \right]\Big].
\label{eq:PM2}
\end{equation}

In this paper, we analyze the uplink performance of a LoRa network based on 
the derivations of \eqref{eq:PM1} and \eqref{eq:PM2} for the aforementioned 
network configuration, while considering three interference 
situations; 1)
 dominant co-SF interferer--takes into account the interfering power of only the dominant 
co-SF interfering device, 2) cumulative co-SF 
interference--considers the impact of cumulative interference from co-SF devices, and 3)
 cumulative co-SF and inter-SF interference--assumes the joint cumulative 
interference from co-SF and inter-SF devices.

%%%%%%%%%%%%%%%%%%%%%%%%%%%%%%%%%%%%%%%%%%%%%%%%%%%%%%%%%%%%%%%%%%%%%%%%%%%
%%%%%%%%%%%%%%%%%%%%%%%%%%%%%%%%%%%%%%%%%%%%%%%%%%%%%%%%%%%%%%%%%%%%%%%%%%%
%%%%%%%%%%%%%%%%%%%%%%%%%%%%%%%%%%%%%%%%%%%%%%%%%%%%%%%%%%%%%%%%%%%%%%%%%%%
%%%%%%%%%%%%%%%%%%%%%%%%%%%%%%%%%%%%%%%%%%%%%%%%%%%%%%%%%%%%%%%%%%%%%%%%%%% 
\section{Interference Free Uplink Performance}
\label{sec:OutageNoInterference}
%%%%%%%%%%%%%%%%%%%%%%%%%%%%%%%%%%%%%%%%%%%%%%%%%%%%%%%%%%%%%%%%%%%%%%%%%%%
%%%%%%%%%%%%%%%%%%%%%%%%%%%%%%%%%%%%%%%%%%%%%%%%%%%%%%%%%%%%%%%%%%%%%%%%%%%
%%%%%%%%%%%%%%%%%%%%%%%%%%%%%%%%%%%%%%%%%%%%%%%%%%%%%%%%%%%%%%%%%%%%%%%%%%%
%%%%%%%%%%%%%%%%%%%%%%%%%%%%%%%%%%%%%%%%%%%%%%%%%%%%%%%%%%%%%%%%%%%%%%%%%%%
When considering an interference-free scenario, uplink outage of the useful signal 
occurs if the SNR of the received signal falls below the 
threshold, $\theta_{\textrm{SF}}$, which depends on the used SF. 
Let $H :=|h|^2$ be the channel gain between the device and the gateway, which is an 
exponential random variable with unit mean i.e., $H \sim \exp(1)$. Then, the   
instantaneous SNR can be defined from \eqref{eq:signalModel} as $\textrm{SNR} = p_t H l(x_{1}) / \sigma^2$ 
where $p_t$ is the transmit power of the node. The success probability, as a 
complement of outage probability, of a device located at distance $x_{1}$ 
from the gateway is given by
\begin{equation}
P_{\textrm{SNR}}(x_{1}, \theta_{\textrm{SF}}) = \mathbb{P}\left[H \ge \frac{
\sigma^2 \theta_{\textrm{SF}}}{p_t 
l(x_{1})}\right] = \exp\left(\frac{\sigma^2 \theta_{\textrm{SF}}}{p_t l(x_{1})
}\right).
\label{eq:outage1}
\end{equation}

Note that \eqref{eq:outage1} is independent of the device intensities $\lambda$ and $
\lambda_m$, and the threshold $\theta_{\textrm{SF}}$ remains fixed in an annulus.

%%%%%%%%%%%%%%%%%%%%%%%%%%%%%%%%%%%%%%%%%%%%%%%%%%%%%%%%%%%%%%%%%%%%%%%%%%%
%%%%%%%%%%%%%%%%%%%%%%%%%%%%%%%%%%%%%%%%%%%%%%%%%%%%%%%%%%%%%%%%%%%%%%%%%%%
%%%%%%%%%%%%%%%%%%%%%%%%%%%%%%%%%%%%%%%%%%%%%%%%%%%%%%%%%%%%%%%%%%%%%%%%%%%
%%%%%%%%%%%%%%%%%%%%%%%%%%%%%%%%%%%%%%%%%%%%%%%%%%%%%%%%%%%%%%%%%%%%%%%%%%%
\section{Uplink Performance Analysis in Poisson Field of Interferers}
\label{sec:OutageInterference}
%%%%%%%%%%%%%%%%%%%%%%%%%%%%%%%%%%%%%%%%%%%%%%%%%%%%%%%%%%%%%%%%%%%%%%%%%%%
%%%%%%%%%%%%%%%%%%%%%%%%%%%%%%%%%%%%%%%%%%%%%%%%%%%%%%%%%%%%%%%%%%%%%%%%%%%
%%%%%%%%%%%%%%%%%%%%%%%%%%%%%%%%%%%%%%%%%%%%%%%%%%%%%%%%%%%%%%%%%%%%%%%%%%%
%%%%%%%%%%%%%%%%%%%%%%%%%%%%%%%%%%%%%%%%%%%%%%%%%%%%%%%%%%%%%%%%%%%%%%%%%%%
In order to analyze the uplink performance in the presence of concurrent 
transmissions, we utilize the stochastic geometry which is a powerful tool to 
model stochastic behavior of a network \cite{haenggi2009interference}. 
Stochastic geometry is based on finding the stochastic average 
of the interference power by summing over the interfering transmissions in 
the network.   
%Stochastic geometry is based on utilizing the stochastic average values for 
%the interference power, which is a summation over the interfering 
%transmitters in the network. All the calculations below are based on [18], [19
%], [24], [25], [30].
To this end, modeling the interference as a shot noise process is widely used 
(e.g.,\cite{haenggi2009interference}), considering the noise components as 
Poisson distributed time instants. For a spatial random process, the time 
instants are replaced with spatial locations of the nodes while the impulse 
responses associated to the time instants are replaced with the path loss model. 

%Then we can assume that (6) describes the interference as a shot noise 
%process. This is a very handy assumption, since it allows the usage of the 
%well known theories to analyze the aggregate interference in the network.

% change based on reviewer 2
%%%When the devices within an annulus are transmitting with the same SF, it 
%%%makes an interesting case for relating the effect of power from 
%%%the strongest interferer to that of the total interference power within the annulus of the desired 
%%%transmitter. To analyze the success 
%%%probability under the strongest interferer, the extreme order statistics 
%%%\cite{kotz2000extreme} can be used.

When the duty cycle constrained devices transmit using the same SF, the 
concurrently active devices within in annulus are reduced significantly. In 
this respect, it is interesting to relate the effect of interfering power from the 
dominant interferer to that of the total interference on the success probability of a 
desired uplink transmission. The success probability under the dominant 
interferer can be analyzed based on the extreme order 
statistics~\cite{kotz2000extreme}.

In what follows, we derive the success probabilities considering 
the dominant co-SF interferer which is built upon \cite{georgiou2017low} for 
the considered path loss model, as well as
the joint impact of cumulative co-SF and inter-SF interference.
%However in case of 
%dominant interferer, the network is assumed to be interference limited only 
%for simplification. This assumption is also made in \cite{georgiou2017low}, 
%and we validate this assumption based on the outage analysis for aggregate 
%interference which reveals the conditions under which this LoRa network 
%setting is truly interference limited and also noise dependent. 

\subsection{SIR Success - Dominant Interferer}
%\subsection{Mathematical Treatment}
%\begin{figure*}[!hb]
%\normalsize
%\hrulefill
%\begin{align} 
%P_{\textrm{SIR}}^*(x_{1}, \delta) \!\geq\! e^{-v} - \frac{\pi ve^{-v} \kappa^\frac{2}{\eta}}{a_i} \left[\frac{l_{x_1}l_x^{\frac{2}{\eta}}}{l_{x_1} - \delta l_x} + \frac{\delta l_{x_1} \left(\eta -2\right) + 2 \delta \left(l_{x_1} + \delta l_x\right){}_2F_1\left(1, \frac{\eta - 2}{\eta}, 2-\frac{2}{\eta}, -\frac{\delta l_x}{l_{x_1}}\right)}{\left(l_{x_1} - \delta l_x\right)\left(\eta-2\right)l_{x_1} l_x^{\frac{2}{\eta} -1} }\right]_{x=\ell_{i}}^{x=\ell_{i-1}} 
%\label{eq:approx}
%\end{align}
%\end{figure*}

We consider the success probability of the desired signal under 
interfering signals of the same SF, however under the effect 
of the strongest interferer only. Let us define the strongest interferer 
$k^*$, in $i$th annulus, as
\begin{equation}
k^* = \underset{x_k\in\Phi_{m,i}\setminus x_1} {\arg \max} \Big\{\mathds{1}_k^{\textrm{SF}_p}p_t G_k l(x_k)
\Big\}, 
\label{eq:Imax} 
\end{equation}
then success probability is determined by the condition that the desired 
signal is $\delta$ times stronger than the dominant interferer      
\begin{equation}
P_{\textrm{SIR}}^*(x_{1}, \delta) = \mathbb{E}_H \left[\mathbb{P}\left[G_{k^*} l(x_{k^*}) \le \frac{
H l(x_{1})}{\delta}\right]\right].
\label{eq:Imax_Outage}
\end{equation}

Let $X_{k^*} = G_{k^*} l(x_{k^*})$, then the success probability under the strongest 
interferer can be calculated from the order statistics. The CDF of the maximum interferer according to the extreme order statistics is $F_{X_{k^*}}(z) = [F_{X_{i}}(z)]^M$, 
where $M$ is a Poisson distributed random variable with mean $v_i = \alpha 
\lambda a_ i$, which is the expected number of concurrently transmitting 
interferers from an annulus of area $a_i$. If $\mathbb{P}(M=m), m=\{0,1,2, \cdots\}$ is 
the probability mass function of $M$, then from the total probability theorem, the order statistics 
in a sample of random size determines $F_{X_{k^*}}(z)$ as
\begin{align} 
F_{X_{k^*}}(x) =  \sum_{k=0}^{\infty} \frac{v^k e^{-v}}{k!}  [F_{X_{i}}(x)]^k.
\label{eq:max_I_CDF}
\end{align}
By using the series representation $e^x =\sum_{k=0}^{\infty} {x^k}/{k!}$, and 
taking expectation over channel gain $H$, the success probability from 
\eqref{eq:Imax_Outage} can be determined as
\begin{align} 
P_{\textrm{SIR}}^*(x_{1}, \delta) =  e^{v_i} \int_0^{\infty}\!\! \exp\bigg(v_i F_{X_i}\Big(
{zl(x_{1})}/{\delta}\Big)\bigg)e^{-z} \mathrm{d}z,
\label{eq:max_I_OutageFinal}
\end{align}
where $F_{X_{i}}(z)$ is the CDF of product distribution \cite{rohatgi2015} of 
the probability density function (PDF) of $l(x_i)$ and $G_i \sim \exp(1)$. 
The distance distribution of the devices to the gateway in $i$th annulus with 
boundaries $[\ell_{i-1}, \ell_{i})$ and area $a_i$  is $2 \pi x/a_i$. Thus, 
for the considered path loss model, the probability density function (PDF) of 
$l(x_i)$ is $f_{l(x_i)}(x) = 2\pi{\kappa^{\frac{2}{\eta}}x^{-\frac{\eta+2}{
\eta}}}/({\eta a_i})$ defined over $l(\ell_{i})\leq x\leq l(\ell_{i-1})$, and $F_{X_{i}}(z)$ is
%The correction of Eq. (8) in \cite{georgiou2017low} is
%
%\begin{align} 
%f_{X_i}(z) \!=\! \frac{2\pi{\kappa^{\frac{2}{\eta}}}}{\eta a_i z^{\frac{\eta+2
%}{\eta}}}\left[\Gamma\! \left({1+\frac{2}{\eta},\frac {z}{l(x)}}\right)\right]
%_{\!x=\ell_{i}}^{\!x=\ell_{i-1}}\!\!, z\in\mathbb{R}_{>0}
%\end{align}
%where $\Gamma(\cdot, \cdot)$ is the upper incomplete gamma function \cite[p.~
%908]{gradshteyn2014table}, and it leads to the cumulative distribution 
%function (CDF) of $X_i$
\begin{align} 
F_{\!X_{i}}\!(z) \!=\! \frac {\pi{\kappa^{\frac{2}{\eta}}}}{a_i}\! \left
[{\frac {{e^{\frac {-z}{l(x)}}\!-\!1}}{l(x)^{\frac {2}{\eta }}}-z^{\!-
\frac {2}{\eta }}\Gamma \left(\!{1\!+\!\frac{2}{\eta},\frac {z}{l(x)}}\right)}\right]_{\!x=\ell_{i}}^{\!x=\ell_{i-1}}\!\!\!,
\end{align}
where $\Gamma(\cdot, \cdot)$ is the upper incomplete gamma function 
\cite{gradshteyn2014table}.

\subsection{SIR Success - Cumulative Interference }
%%%%%%%%%%%%%%%%%%%%%%%%%%%%%%%%%%%%%%%%%%%%%%%%%%
In this section, we perform the SIR 
based outage or success probability analysis of the desired signal subjected to aggregate co- and 
inter-SF interference. The objective is to find how cumulative interference, 
as apposed to the dominant interferer, affects the performance considering 
the dominant interferer analysis to be an upper bound on the
outage. 

\subsubsection{SIR Success under Co-Spreading Factor Interference}

The SIR experienced by the desired signal, from the $i$th annulus, at the gateway 
under concurrent co-SF interference~is
\begin{equation}
\textrm{SIR}_{\textrm{CSF}} = \frac{H l(x_{1})}{\underbrace{\!\!\!\!\!\sum_{x_
k\in\Phi_{m,i}\setminus x_1}\!\!\! {\mathds{1}_k^{\textrm{SF}_p} G_k l(x_k)}}_{
\mathcal{I}_
{\textrm{CSF}}}}.
\vspace{-4pt}
\label{eq:SINR_1}
\end{equation}
Let $\delta_{ii}{[\textrm{dB}]}$ be the threshold needed for the desired signal 
at SF$_i$ against interferers using the same SF$_i$, then the condition 
for success probability $\mathbb{P}\left[\textrm{SIR}_{\textrm{CSF}}  \ge 
\delta_{ii}\right]$ can be expanded as  
\begin{align}
P_{\textrm{SIR}}(x_{1}, \delta_{ii}) & = \mathbb{E}_{\mathcal{I}_{\textrm{CSF
}}}\left[\mathbb{P}\left[H \ge \frac{\delta_{ii} 
\mathcal{I}_{\textrm{CSF}}}{p_t l(x_{1})}\right]\right] \nonumber \\
							 & \stackrel{(a)}{=} \mathbb{E}_{\mathcal{I}_{\textrm{CSF}}}
\left[\exp\left(-\frac{\delta_{ii} 
\mathcal{I}_{\mathrm{CSF}}}{p_t l(x_{1})}\right)\right] \nonumber \\
							 & \stackrel{(b)}{=} \mathcal{L}_{\mathcal{I}_{\textrm{CSF}}}
\left(\frac{\delta_{ii}}{p_t l(x_{1})}
\right),
\label{eq:outage2}
\end{align}
where (a) in \eqref{eq:outage2} follows from $H\sim \exp(1)$. Also in (b), $
\mathcal{L}_{\mathcal{I}_{\textrm{CSF}}}\left(\cdot\right)$ is the 
Laplace transform (LT) at $s = {\delta_{ii}}/{\left(p_t l(x_{1 }\!)\right)}$ 
of the cumulative interference power $\mathcal{I}_{\mathrm {CSF}}$, which 
can be evaluated with the probability 
generating functionals (PGFLs). 
Using the LT definition, from \eqref{eq:SINR_1} and \eqref{eq:outage2} we have 
\begin{align}
\mathcal{L}_{\mathcal{I}_{\mathrm{CSF}}}(s)\!& = \mathbb{E}\left[\exp\left(-s \mathcal{I}_{\mathrm{CSF}}\right)\right] \nonumber \\
& = \mathbb{E}_{\Phi_{m,i}, G_k}\!\!\left[\exp\left(\!-s\!\!\!\!\!\sum_{x_k\in \Phi_{m,i} \setminus x_{1}}{\!\!p_t G_k l(x_k)}\!\right)\!\right] \nonumber \\
& = \mathbb{E}_{\Phi_{m,i}, G_k}\!\!\left[\prod_{x_k\in \Phi_{m,i} \setminus x_{1}}{\!\!\!\!\!\!\exp\Big(-s{p_t G_k l(x_k)}\!\Big)}\!\right],
\label{eq:Laplace_I_CSF1}
\end{align}
%\mathbb{E}_{\Phi}\left[\prod \mathbb{E}_G\left[\exp(-s \beta G l(d_j))\right]\right]
where the expectation is over the point process $\Phi_{m,i}$ and channel 
gain $G$. The expectation with respect to $G$ can be moved inside, as $G$ is 
independent from the PPP. Then, using the moment generation function of exponential random variable with $G\sim \exp\left(1\right)$ in \eqref{eq:Laplace_I_CSF1} yields
\begin{equation}
\mathcal{L}_{\mathcal{I}_{\mathrm{CSF}}}(s) = \mathbb{E}_{\Phi_{m,i}}\left[\prod_{x_k\in \Phi_{m,i} \setminus x_{1}}\frac{1}{1+s p_t l(x_k)}\right].
\label{eq:Laplace_I_CSF11}
\end{equation}

By using the PGFL of homogeneous PPP \cite{haenggi2009interference} with 
respect to the inner function of \eqref{eq:Laplace_I_CSF11} i.e., $\mathbb{E}\left[\prod_i{\mathcal{U}\left(x_i\right)}\right] = \exp\left(-\lambda\int_{\mathbb{R}^2}{1-\mathcal{U}\left(x\right)}dx\right)$ for any no-negative function $\mathcal{U}\left(x_i\right)$, we have
%\begin{align}
%\mathcal{L}_{I_{\mathrm{CSF}}}(s)  & =  \exp \bigg(\!-2\pi\beta\lambda \times \nonumber \\
%& \int_{\ell_i}^{\ell_{i+1}^{-}}{\!\!\!\left(\!1 - \mathbb{E}_G\left[\exp\left(\!-\frac{s G l(x)}{l(x_1)}\right)\right]\right)x\textrm{d}x\!\bigg)}
%\label{eq:Laplace_I_CSF2}
%\end{align}
\begin{equation}
\mathcal{L}_{\mathcal{I}_{\mathrm{CSF}}}(s) \! = \! \exp \bigg(\!\!\!-\alpha\lambda \!\int_{\mathbb{R}^2}{\!\left(\!1 - \frac{1}{1+s p_t l(x_j)}
\right)\textrm{d}x\!\bigg)}.
\label{eq:Laplace_I_CSF2}
\end{equation}
Finally, the transformation of Cartesian to polar coordinates $x_j = (x,\theta
)$ gives
\begin{equation}
\mathcal{L}_{I_{\mathrm{CSF}}}(s)  = \exp \left(-2\pi\alpha\lambda \int_{\ell_
{i-1}}^{\ell_{i}^{-}}{\frac{s p_t l(x)}{1 + s p_t l(x)}x\textrm{d}x}\right),
\label{eq:Laplace_I_CSF3}
\end{equation}
where $\ell_{i-1}$ and $\ell_{i}^-$ are the inner and outer boundaries 
of the $i$th annulus. Replacing $s$ gives the success probability of a 
zero-noise network i.e., for an interference-limited case. 

Using \eqref{eq:Laplace_I_CSF3} in \eqref{eq:Laplace_I_CSF1}, we get the 
success probability of a packet from a device located at distance $x_{1}$, within an annulus $[\ell_{i-1}, \ell_{i})$,  
from the gateway under same SF interference as
%\vspace{-5pt}
\begin{align}
& P_{\textrm{SIR}}(x_{1}, \delta_{ii}) = \exp \bigg(-2\pi\alpha\lambda \mathcal{I}\left(x_1, \delta_{ii}, \{\ell_{i-1}, \ell_i\}\right)\bigg),
\label{eq:Outage_CSF}
\end{align}
%\vspace{-5pt}
where
%\vspace{-3pt} 
\begin{align}
\mathcal{I}\left(x_1, \delta_{ii}, \{\ell_{i-1}, \ell_i\}\right) =  \int_{\ell_{i-1}}^{\ell_{i}^{-}} \frac{\delta_{ii}l(x)}{l(x_1) + \delta_{ii}l(x)}{x\textrm{d}x}.
\label{eq:Ax}
\end{align}

%%%%%%%%%%%%%%%
\subsubsection{SIR Success under Inter-Spreading Factor Interference}
%%%%%%%%%%%%%%%
When inter-SF interference is considered, the outage occurs when the SIR of 
the desired signal of $\mathrm{SF}_i$ goes below the threshold $\delta_{ij}[
\textrm{dB}]$ (see SIR matrix \eqref{eq:sinrMatrix}) when the concurrent 
transmissions of quasi-orthogonal SFs $\mathrm{SF}_j$ are interfering. Let the 
device transmitting the useful signal be located in the $i$th annulus, then 
concurrent transmissions with orthogonal SFs will originate from $\mathcal{K}
\setminus i$ annuli i.e., all the annuli except $i$th annulus and SIR can be defined as
\begin{equation}
\textrm{SIR}_{\textrm{ISF}} = \frac{H l(x_{1})}{\underbrace{\!\! \sum_{j = 1, 
j \neq i}^{K}\,{\sum_{x_k\in\Phi_{m,j}} \mathds{1}_k^{\textrm{SF}_q}G_k l(x_k)}}_{\mathcal{I}_{\textrm{ISF
}}}}.
\label{eq:SINR_2}
\end{equation} 
However, the desired signal has a different SIR margin with respect to the 
interfering transmissions from each annulus. As all 
annuli are disjoint, from the independence of PPP, the success 
probability of the desired transmission in the $i$th annulus under inter-SF interference, 
originating from $\mathcal{K}\setminus i$ annuli, can be 
determined from \eqref{eq:outage2} as, 
\begin{align} 
P_{\textrm{SIR}}(x_{1}, \delta_{ij}) = \prod_{j=1,j\neq i}^K{\mathcal{L}_{
\mathcal{I}_{\mathrm{ISF}}}\left(\frac{\delta_{ij}}{p_t l(x_1)}\right)},
\label{eq:Outage_ICSF}
\end{align}
where $\mathcal{L}_{\mathcal{I}_{\mathrm{ISF}}}(\cdot)$ is the LT of 
cumulative inter-SF interference power at $s = \delta_{ij} / (p_t l(x_1))$ 
which can be found from \eqref{eq:Laplace_I_CSF1}-\eqref{eq:Laplace_I_CSF3} 
\begin{align} 
P_{\textrm{SIR}}(x_{1}, \delta_{ij})\! =\! \exp\!\bigg(\!\!\!-\!2
\pi\alpha\lambda \!\!\!\!\sum_{j=1, j\neq i}^K\!\!\!\! 
\mathcal{I}\!\left(x_1, \delta_{ij}, \!\{\ell_{j-1}, \ell_j\}\right) \!\!\bigg
),
\label{eq:Laplace_ICSF}
\end{align}
where $\mathcal{I}\left(\cdot\right)$ is defined in \eqref{eq:Ax}.
%which finds the cumulative interference power from deployment region $\mathcal
%{S}\setminus \mathcal{S}_i$, that is excluding $\mathcal{S}_i$- the co-SF 
%interference region. In \eqref{eq:Outage_ICSF}, $\mathcal{I}\left(
%\cdot\right)$ is defined in \eqref{eq:Ax}. 
%\begin{figure}[!ht]
	%\centering
		%\includegraphics[width=0.9\linewidth]{InterferenceModel.png}
	%\caption{Possible inter- and co-spreading factor interference 
%configurations: inter-spreading factor interference is generated from the 
%gray region while blue region marks the co-spreading factor interference.}
	%\label{fig:interModel}
%\end{figure}

%Note that in some studies a different SIR matrix is reported, where a SF$_i$ 
%has different SIR requirement with respect to every other orthogonal SF$_j$. 
%If that's the case the success probability under inter-SF interference is 
%simply found as
%\begin{align} 
%& P_{\textrm{SIR}}(x_{1}, \delta_{ij}) = \!\!\!\prod_{j=1, j\neq i}^K{\!\!\exp\!\!\bigg(\!\!\!-\!2\pi\alpha\lambda\mathcal{I}\!\left(x_1, \!\delta_{ij}, \!\{\ell_{j-1}, \!\ell_j\}\right)\!\!\bigg)}
%\label{eq:Outage_ICSF_difSINR}
%\end{align}

%%%%%%%%%%%%%%%
\subsubsection{SIR outage under Co- and Inter-SF Interference}
%%%%%%%%%%%%%%%
For a device located at $x_1$ in the $i$th annulus, the success probability 
under the co- and inter-SF interference together can be determined from
\eqref{eq:Outage_CSF} and \eqref{eq:Laplace_ICSF} as
\begin{equation}
P_{\textrm{SIR}}^{\Pi}(x_{1}, \delta_{ij})\! =\! \exp\!\bigg(\!\!-\!2
\pi\alpha\lambda \sum_{j=1}^K 
\mathcal{I}\!\left(x_1, \delta_{ij}, \{\ell_{j-1}, \ell_j\}\right) \!\bigg
),
\label{eq:P_SIR_Total}
\end{equation}
%where $\mathcal{L}_{\mathcal{I}_{\textrm{CSF}}}(\cdot)$ and $\mathcal{L}_{
%\mathcal{I}_{\textrm{ISF}}}(\cdot)$ are determined from 
%\eqref{eq:Laplace_I_CSF3} and \eqref{eq:Laplace_ICSF}, respectively.

%On the other hand the SINR based outage probability is determined from 
%\eqref{eq:Outage_CSF} and \eqref{eq:Outage_ICSF}
%\begin{equation}
%P_{\textrm{SINR}}^{\Pi}(x_{1}, \delta_{ii}, \delta_{ij}) = P_{\textrm{SINR}}(x
%_{1}, \delta_{ii}) P_{\textrm{SINR}}(x_{1}, \delta_{ij})
%\label{eq:P_SINR_Total}
%\end{equation}

%%%%%%%%%%%%%%%%%%%%%%%%%%%%%%%%%%%%%%%%%%%%%%%%%%%%%%%%%%%%%%%%%%%%%%%%%
%%%%%%%%%%%%%%%%%%%%%%%%%%%%%%%%%%%%%%%%%%%%%%%%%%%%%%%%%%%%%%%%%%%%%%%%% 
\subsection{Coverage Analysis}
\label{sec:CoverageAnalysis}
%%%%%%%%%%%%%%%%%%%%%%%%%%%%%%%%%%%%%%%%%%%%%%%%%%%%%%%%%%%%%%%%%%%%%%%%%
%%%%%%%%%%%%%%%%%%%%%%%%%%%%%%%%%%%%%%%%%%%%%%%%%%%%%%%%%%%%%%%%%%%%%%%%% 

Based on the analyzed success probabilities, $\mathcal{Y}\in \{P_{\textrm{SNR
}}, P_{\textrm{SIR}}^*, P_{\textrm{SIR}}, P_{\textrm{SIR}}^{\Pi}\}$, from \eqref{eq:PM2} the coverage probability 
can be determined by averaging over the distance distribution of $x_1$, $f_D(x_1)$ as
\begin{equation}
P_c\left[\mathcal{Y}\right] = \int_{d>0}^R{\mathcal{Y}\cdot f_D(x_1)\textrm{d}x_1},
\label{eq:CovProbDef}
\end{equation}

Using the PDF of the distance of a uniformly distributed random devices 
within the area $\pi R^2$, the coverage probability for the assumed geometry 
of the LoRa network, from \eqref{eq:CovProbDef}, is
\begin{equation}
P_c\left[\mathcal{Y}\right] = \frac{2}{R^2}\sum_{i=1}^n\bigg({\int_{\ell_{i-1
}}^{\ell_{i}^-}{\mathcal{Y}\cdot x_{1}\textrm{d}x_{1}}}\bigg).
\label{eq:eq:CovProbMultiAnnuli}
\end{equation}

%%%%%%%%%%%%%%%%%%%%%%%%%%%%%%%%%%%%%%%%%%%%%%%%%%%%%%%%%%%%%%%%%%%%%%%%%%%
%%%%%%%%%%%%%%%%%%%%%%%%%%%%%%%%%%%%%%%%%%%%%%%%%%%%%%%%%%%%%%%%%%%%%%%%%%%
%%%%%%%%%%%%%%%%%%%%%%%%%%%%%%%%%%%%%%%%%%%%%%%%%%%%%%%%%%%%%%%%%%%%%%%%%%%
%%%%%%%%%%%%%%%%%%%%%%%%%%%%%%%%%%%%%%%%%%%%%%%%%%%%%%%%%%%%%%%%%%%%%%%%%%% 
\section{Results and Discussion}
\label{sec:SimulationResults}
%%%%%%%%%%%%%%%%%%%%%%%%%%%%%%%%%%%%%%%%%%%%%%%%%%%%%%%%%%%%%%%%%%%%%%%%%%%
%%%%%%%%%%%%%%%%%%%%%%%%%%%%%%%%%%%%%%%%%%%%%%%%%%%%%%%%%%%%%%%%%%%%%%%%%%%
%%%%%%%%%%%%%%%%%%%%%%%%%%%%%%%%%%%%%%%%%%%%%%%%%%%%%%%%%%%%%%%%%%%%%%%%%%%
%%%%%%%%%%%%%%%%%%%%%%%%%%%%%%%%%%%%%%%%%%%%%%%%%%%%%%%%%%%%%%%%%%%%%%%%%%%

\begin{figure*}[!t] 
    \centering
  \subfloat[$R=6$ km]{%
       \includegraphics[width=0.50\linewidth]{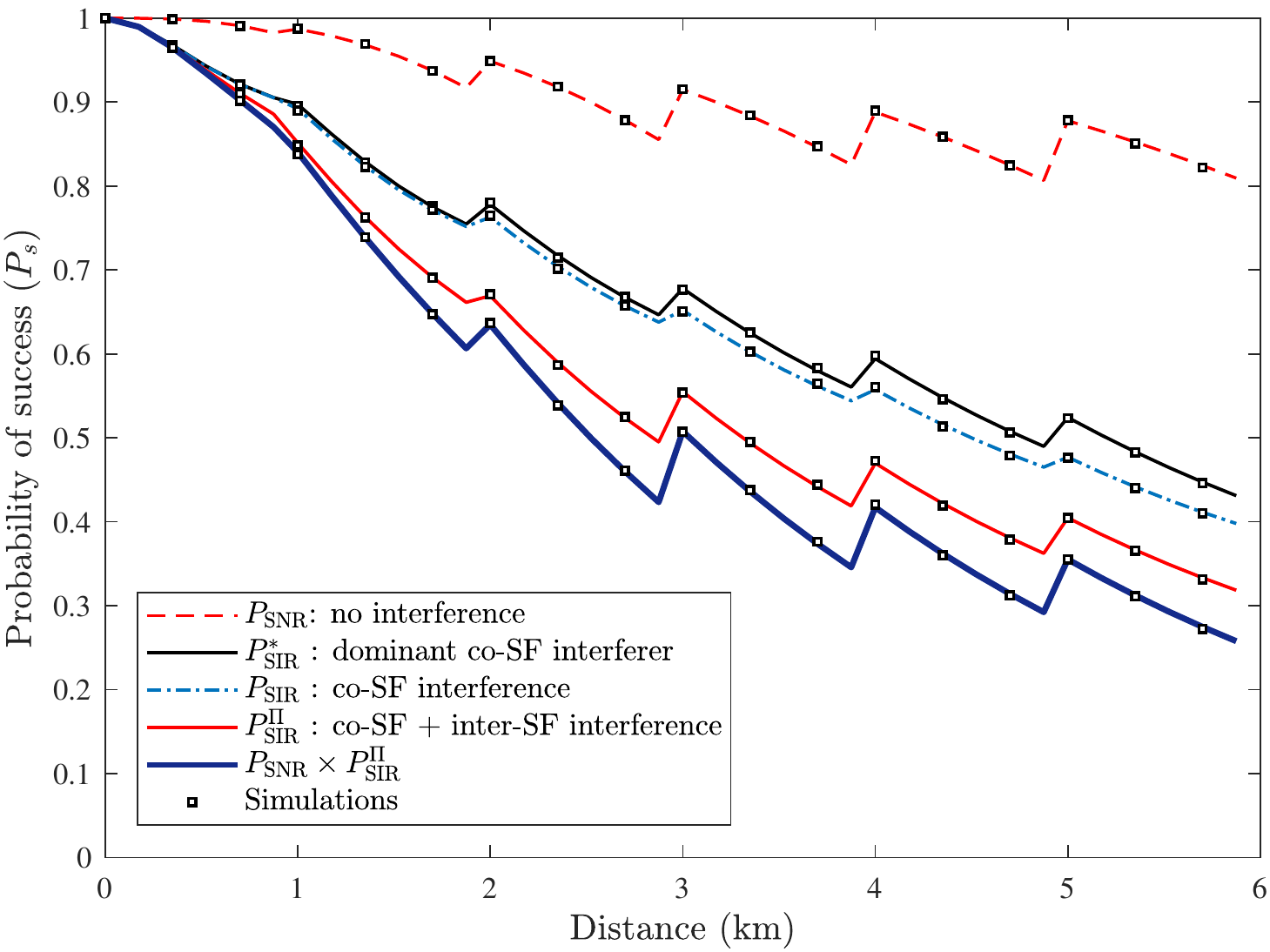}\label{fig:1a}}\hfill
  \subfloat[$R=12$ km]{%
        \includegraphics[width=0.50\linewidth]{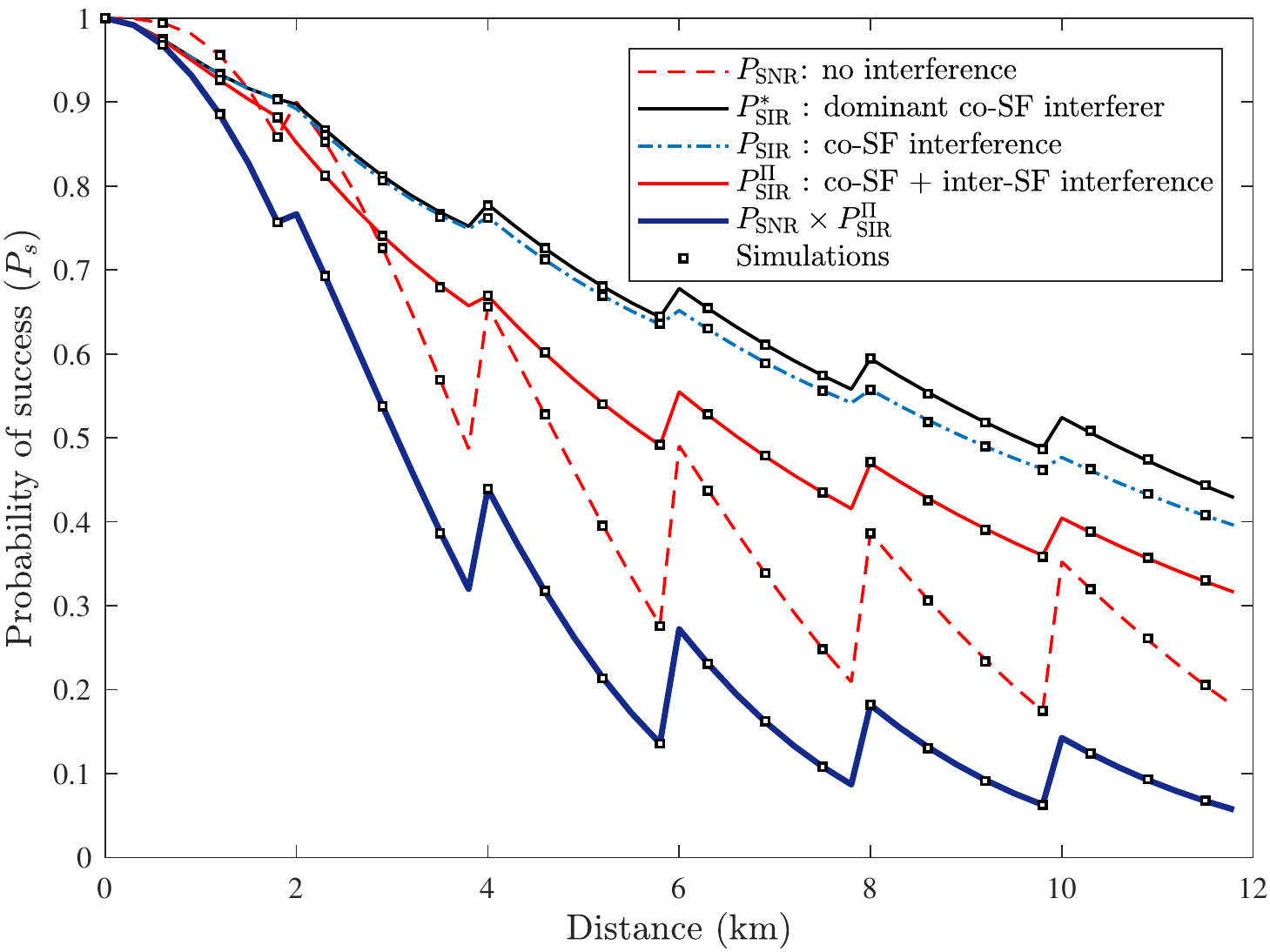}\label{fig:1b}}
  \caption{Outage probability under studied outage conditions with average 
number of end-devices (EDs) $\bar{N}=1500$, and $\eta = 3$. Solid lines  are obtained via 
numerical evaluation of \eqref{eq:max_I_OutageFinal}, \eqref{eq:Outage_CSF} and \eqref{eq:P_SIR_Total} , whereas the markers '\framebox(2.4,2.4){}' represent simulation results.}
\vspace{-12pt}
  \label{fig:OutProb}      
\end{figure*}

%\subsection{Results}
% After review 1
%%%%In this section, we validate the analytical models for success probability 
%%%%and network coverage through Monte Carlo simulation. Each 
%%%%performance point in distance is averaged over $10^5$ 
%%%%realizations of nodes location, which are distributed with homogeneous PPP in 
%%%%a region $\mathcal{S}$ of radius $R$. The matching of the numerical and 
%%%%simulation results demonstrates the accuracy of the developed  
%%%%models. The parameters used to obtain 
%%%%the results are given in Table~\ref{tab:Tab1} while the multi-annuli 
%%%%allocation of SFs is based on Table~\ref{tab:Tab2}, where the 
%%%%SF-dependent SNR thresholds are from \cite{semtech120022}. The SIR 
%%%%thresholds, needed for the success of desired signal under co-SF and inter-SF 
%%%%interference are given in SIR matrix \eqref{eq:sinrMatrix}.

In this section, we validate the analytical models for 
success probability 
and network coverage through Monte Carlo simulations. In the simulation setup, 
the devices are distributed in a disk of radius $R$ according to a 
homogeneous PPP of intensity $\lambda=\bar{N}/A$, where 
$\bar{N}$ is the average number of devices. The disk is 
divided into $K$ disjoint annuli of equal width $R/K$ (unless specified otherwise), where $K$ is 
the number of available SFs. To calculate the success probability of a 
desired uplink transmission at the gateway, we move the node location from 
the inner to the outer radius of an annulus with a minimal fixed step size. At 
each location, the success of the desired transmission is evaluated for two 
conditions: 1) the SNR of the useful signal is above the 
SF-dependent SNR threshold, 2) the SIR of the same signal exceeds the 
SIR threshold under simultaneously active devices causing co-SF and inter-SF 
interference. The set of concurrently transmitting devices is determined by 
the duty cycle parameter $\alpha$. Then, we find the success probability at 
each distance as expected frequency over $10^5$ independent 
realizations of nodes' distribution. Similarly, we find the coverage 
probability with respect to $\bar{N}$ under noise and the considered interference 
conditions.

The matching of the numerical and 
simulation results demonstrates the accuracy of the developed  
models. The parameters used to obtain 
the results are given in Table~\ref{tab:Tab1} while the allocation of SFs in 
a single-cell multi-annuli network is based on Table~\ref{tab:Tab2}, where the 
SF-dependent SNR thresholds are from \cite{semtech120022}. The SIR 
thresholds
%, needed for the success of desired signal under co-SF and inter-SF 
%interference 
are given in SIR matrix \eqref{eq:sinrMatrix}.

%%%%%% OUTAGE PROPBABILITY FIGURES%%%%%%%%%%
\subsection{Success Probability}
In Fig.~\ref{fig:OutProb}, the success probability of the assumed LoRa 
network for the studied SNR- and SIR-based conditions is shown for two 
different cell sizes; $R =6$ km in Fig.~\ref{fig:OutProb}(a) and $R = 12$ km 
in Fig.~\ref{fig:OutProb}(b). The $P_{\textrm{SNR}}$ curve from 
\eqref{eq:outage1} shows that the success probability decreases with respect 
to the distance from the gateway due to path loss and fading. 
It is however interesting to note how success probability 
improves at the annuli transitions (e.g., at $x_1 \in \{1, 2, \cdots, 5\}$ km for $R=6$ 
km) with the use of higher SFs. The gain in the performance is due 
to the lower receiver sensitivities and hence the lower required SNRs for 
higher 
SFs. As a result, a higher SF yields a positive performance jump at the 
inner boundary of each annulus and the multi-annuli allocation of SFs 
gives a saw-tooth trend to the interference-free success probability. Note 
that the 
gain depends on the SF-dependent SNR threshold $\theta_{\textrm{SF}}$ and on the strategy to allocate
SFs. In these results, we assume an equal-interval based annuli structure as described in Sec.~\ref{subsec:SysGeom}, where the EDs in the inner most annulus use the lowest SF and the SFs' strength increases for the outer annuli.  

When it comes to study the network performance under self-interference, we 
use three SIR based metrics: strongest co-SF interferer 
($P_{\textrm{SIR}}^*$), aggregate co-SF interference ($P_{\textrm{SIR}}$) and 
aggreagte co-SF and inter-SF interference ($P_{\textrm{SIR}}^{\Pi}$). The 
respective curves obtained from \eqref{eq:max_I_OutageFinal}, 
\eqref{eq:Laplace_I_CSF3} and \eqref{eq:Laplace_ICSF} are shown in~Fig.~\ref{fig:OutProb}. From these results, we can make the following remarks:
\bgroup
\def\arraystretch{1.2}
%  1 is the default, change whatever you need
%\setlength{\tabcolsep}{pt}
\begin{table}[!t]
%\medium
\centering
 \caption{Parameters According to LoRaWAN Specifications}
\label{tab:Tab1}
\centering
\begin{tabular}{lll}
\noalign{\hrule height 1pt}
\textbf{Parameter}  & \textbf{Symbol} &\textbf{Value}\\
\noalign{\hrule height 1pt}
Signal bandwidth    & $B$   	 & $125$ kHz\\
Carrier frequency   & $f_c$ 	 & $868.10$ MHz \\
Noise power density & $N_0$ 	 & $-174$ dBm/Hz\\
Noise figure        &  NF  	 	 & $6$ dB \\
Transmit power      & $p_t$ 	 & $14$ dBm \\
Duty cycle 	  			& $\alpha$ & $0.33$\% \footnotemark\\
Pathloss exponent   & $\eta$   & $3$ \\
\noalign{\hrule height 1pt}
\vspace{-18pt}
\end{tabular}
\end {table}
\egroup
\bgroup
\def\arraystretch{1.2}%  1 is the default, change whatever you need
\begin{table}[!t]
%\medium
\centering
 \caption{LoRa Spreading Factor Specific Characteristics}
\label{tab:Tab2}
\centering
\begin{tabular}{cccc}
\noalign{\hrule height 1pt}
\textbf{Annulus} & \textbf{SF} & \textbf{SNR thresh. ($\theta$)}  & \textbf{Range}\\
 & & (dB) & (m)\\
\noalign{\hrule height 1pt}
1 & 7  & $-6$   & $\ell_0 - \ell_1$ \\
2 & 8  & $-9$   & $\ell_1 - \ell_2$ \\
3 & 9  & $-12$  & $\ell_2 - \ell_3$ \\
4 & 10 & $-15$  & $\ell_3 - \ell_4$ \\
5 & 11 & $-17.5$ & $\ell_4 - \ell_5$\\
6 & 12 & $-20$   & $\ell_5 - R$\\
\noalign{\hrule height 1pt}
\vspace{-18pt}
\end{tabular}
\end {table}
\egroup
\footnotetext{Note that as per ERC Recommendation 70-03 \cite{ERCrecom2017}, 
the duty cycle limitation of $<1$\% is on the whole h1.4 frequency band of EU 
868.0 - 868.6 MHz. Therefore for signal bandwidth of 125 kHz, there can be 3 
channels in total and it is per-channel duty cycle limitation.}
\begin{figure*}[!t] 
    \centering
  \subfloat[$R=6$ km]{%
       \includegraphics[width=0.50\linewidth]{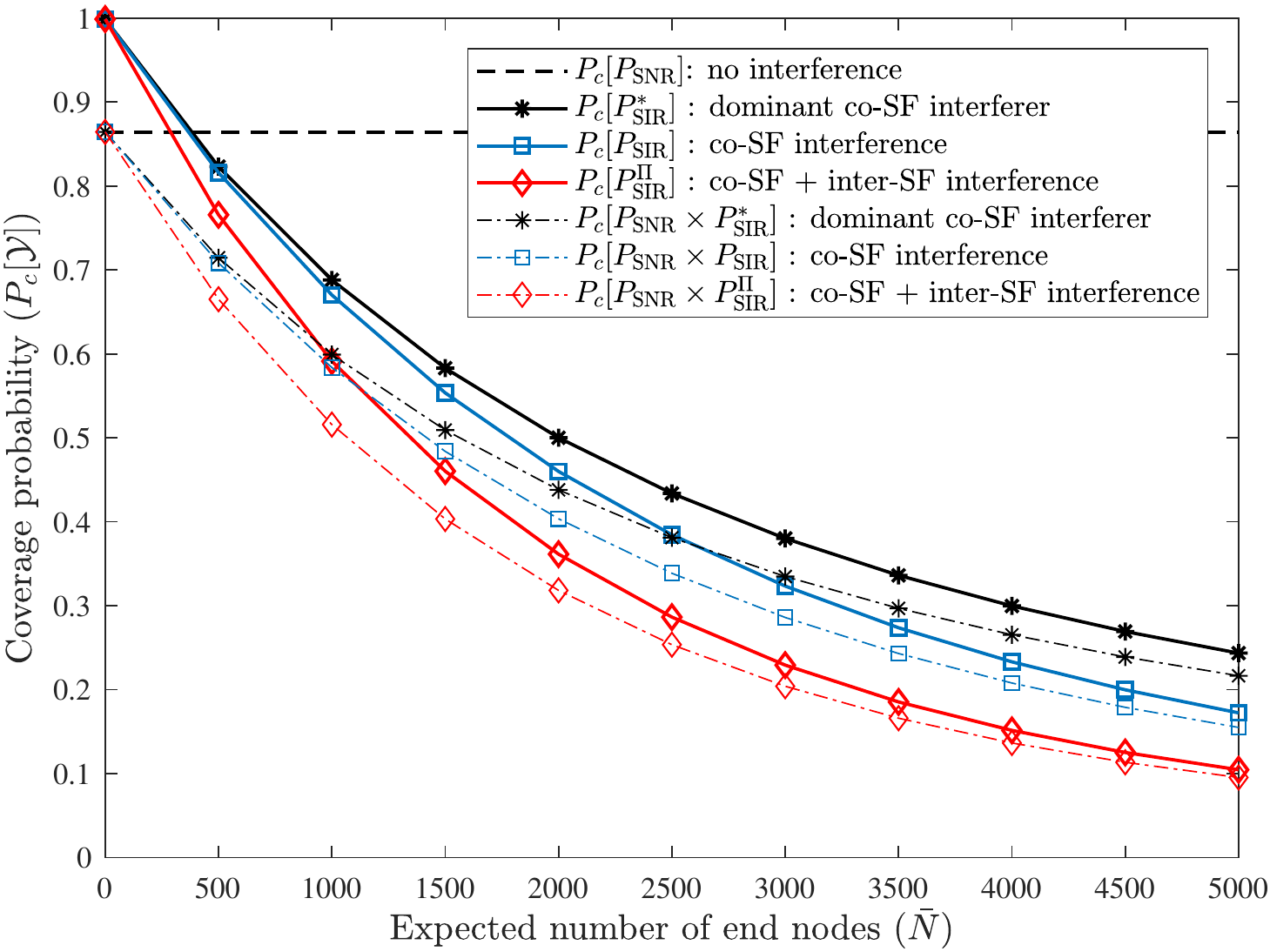}\label{fig:2a}}\hfill
  \subfloat[$R=12$ km]{%
        \includegraphics[width=0.50\linewidth]{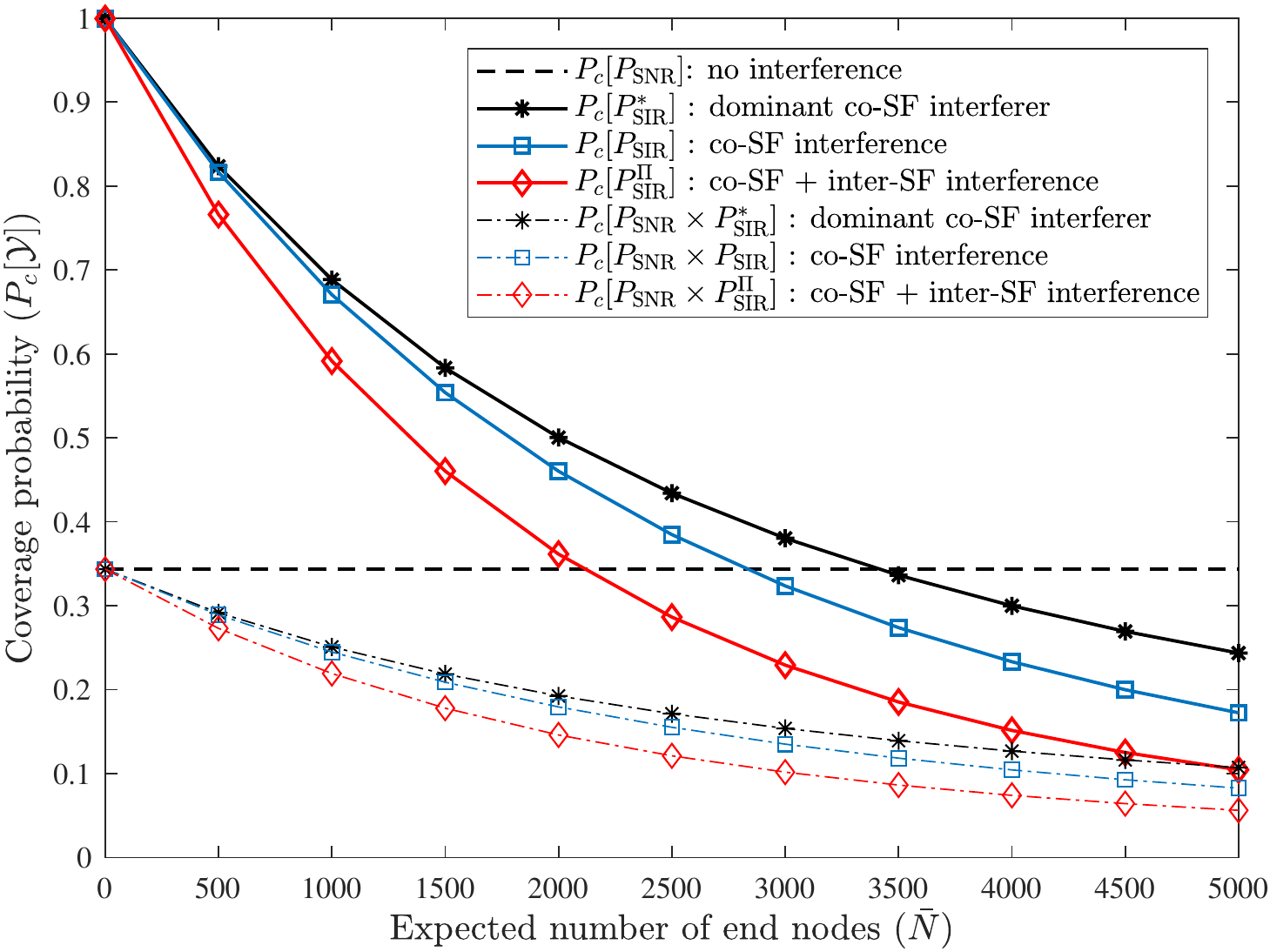}\label{fig:2b}}	
  \caption{Coverage probability with respect to average number of EDs for the studied outage conditions at $\eta = 3$ and two different cell sizes.}
	\vspace{-12pt}
  \label{fig:CovProb} 
\end{figure*}
\begin{itemize}[leftmargin=*]
\item In general, the saw-tooth trend in the success probability curves, produced by 
the ED switching to a higher SF region, is observed in all the considered 
interference conditions, although less prominent compared to non-interference case. 
This small gain can be attributed to the relative location of the 
ED at the inner boundary of the annulus compared to the co-SF interfering 
EDs, thus making it more probable for the ED to achieve an SIR of at 
least 1 dB. Despite the small gain, the 
success probability decreases with the distance from the gateway due to 
increase in number of EDs causing co-SF and/or inter-SF interference. 

\item At annuli boundaries, the small gain in 
$P_{\textrm{SIR}}^*$, $P_{\textrm{SIR}}$ and $P_{\textrm{SIR}}^{\Pi}$ is 
determined by the co-SF and/or inter-SF based SIR thresholds in addition to the SF allocation strategy. 

\item The success probability under 
cumulative co-SF interference ($P_{\textrm{SIR}}$), modeled as a shot-noise 
process, follows the success probability obtained under dominant interferer 
($P_{\textrm{SIR}}^*$). In fact, $P_{\textrm{SIR}}^*$ serves as an upper 
bound but it loses its tightness for higher SFs annuli because the 
number of EDs is proportional to the area $a_i$ of an annulus and hence the 
impact of cumulative interference increases. 

\item The inter-SF interference can cause up to  15\% loss in the success 
probability ($P_{\textrm{SIR}}^{\Pi}$) as compared to co-SF interference 
only, which is significant enough to be taken into consideration in a 
realistic scalability analysis.
\end{itemize}

The SNR- and SIR-based performance indicators must be observed together to 
understand the dominant cause of the performance degradation. Although in 
very dense deployment scenarios, the interference can be a main cause of drop 
in the performance. However considering the cell size, limited transmit power 
of EDs and signal propagation conditions, the impact of background noise must 
also be accounted for. As can be observed from Fig.~\ref{fig:OutProb}, the 
impact of interference on the success probability dominates for a cell of 
radius $R=6$ km (see Fig.~\ref{fig:OutProb}\subref{fig:1a}) while for $R=12$ 
km the impact of noise is more than the 
interference~(see Fig.~\ref{fig:OutProb}\subref{fig:1b}).
In essence, as $\bar{N}$ is the same for both cell sizes, we observe that the relative impact of the interference on the desired transmission remains the same. However, the success probability under noise is degraded more in 
Fig.~\ref{fig:OutProb}\subref{fig:1b} due to a bigger cell size. 
The overall impact can be deduced based on the joint success probability by 
multiplying $P_{\textrm{SNR}}$ with $P_{\textrm{
SIR}}^{\Pi}$. 

\subsection{Coverage Probability}
%%%%%% COVERAGE PROPBABILITY FIGURES%%%%%%%%%%

\begin{figure*}[!ht] 
    \centering
  \subfloat[$\alpha = 0.33\%$, $p_t = 14$ dBm]{
       \includegraphics[width=0.33\linewidth]{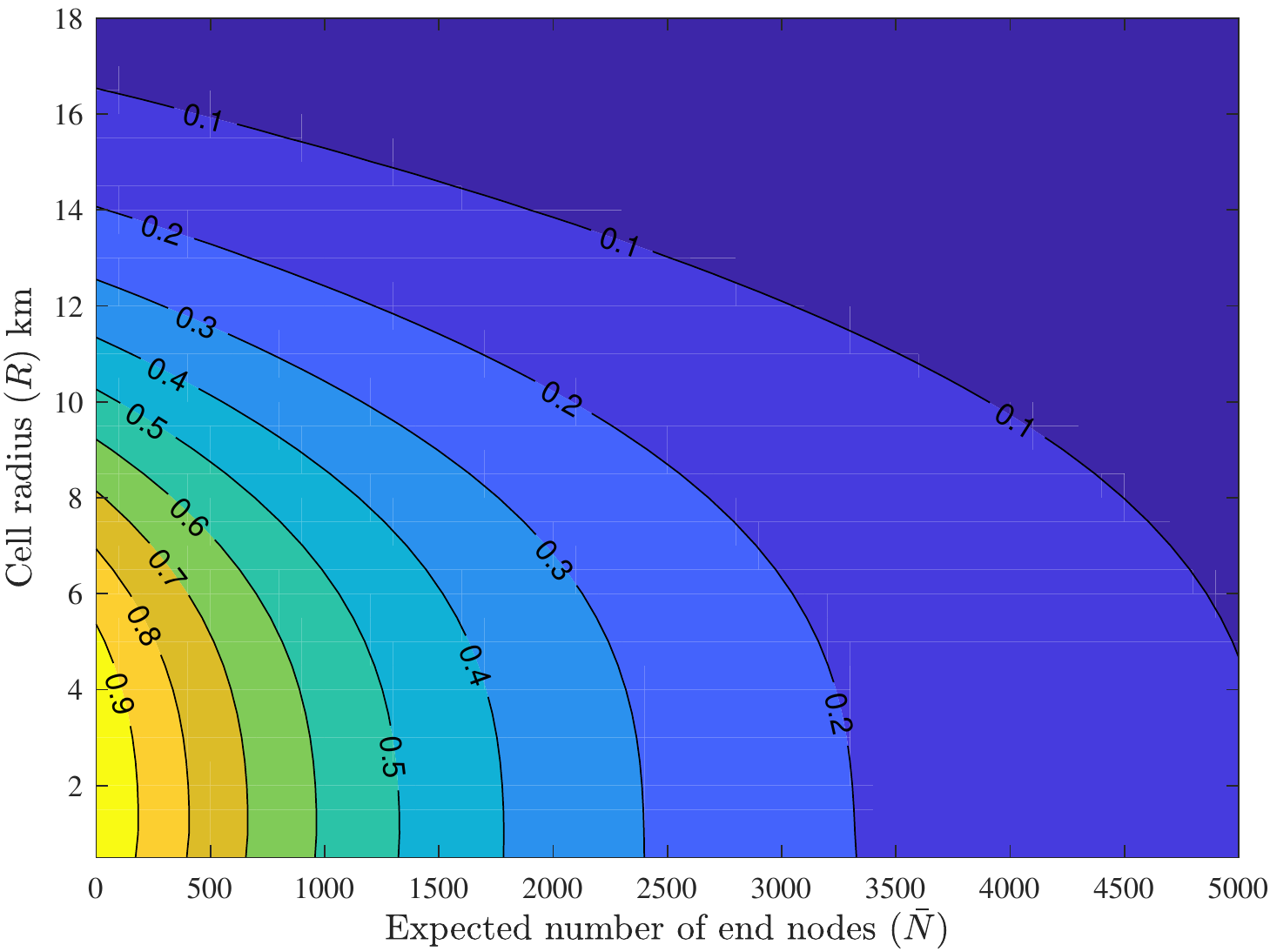}\label{fig:3a}}\hfill
  \subfloat[$\alpha = 0.05\%$, $p_t = 14$ dBm]{%
        \includegraphics[width=0.33\linewidth]{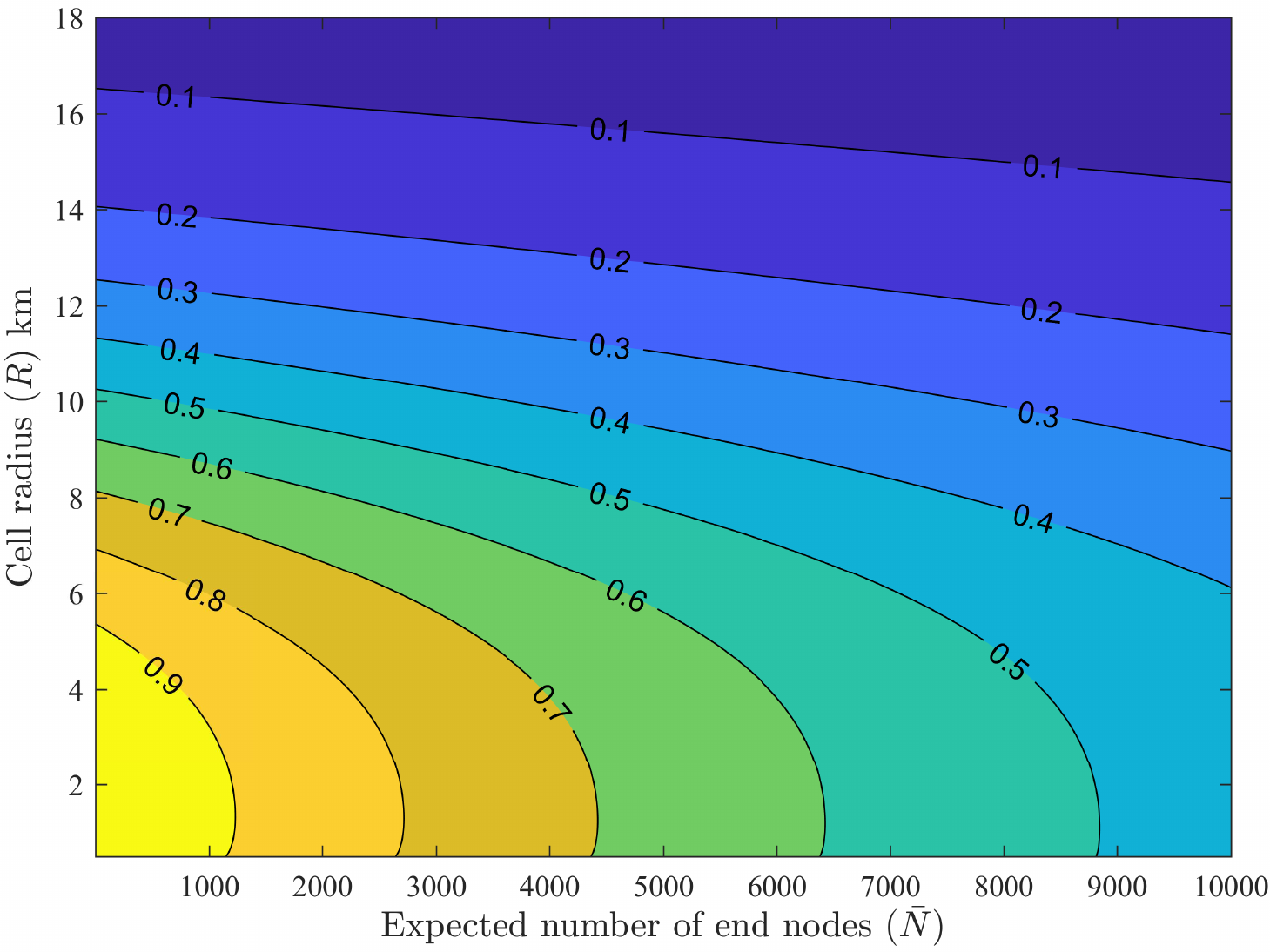}\label{fig:3b}}\hfill	
  \subfloat[$\alpha = 10\%$, $p_t = 27$ dBm]{%
        \includegraphics[width=0.33\linewidth]{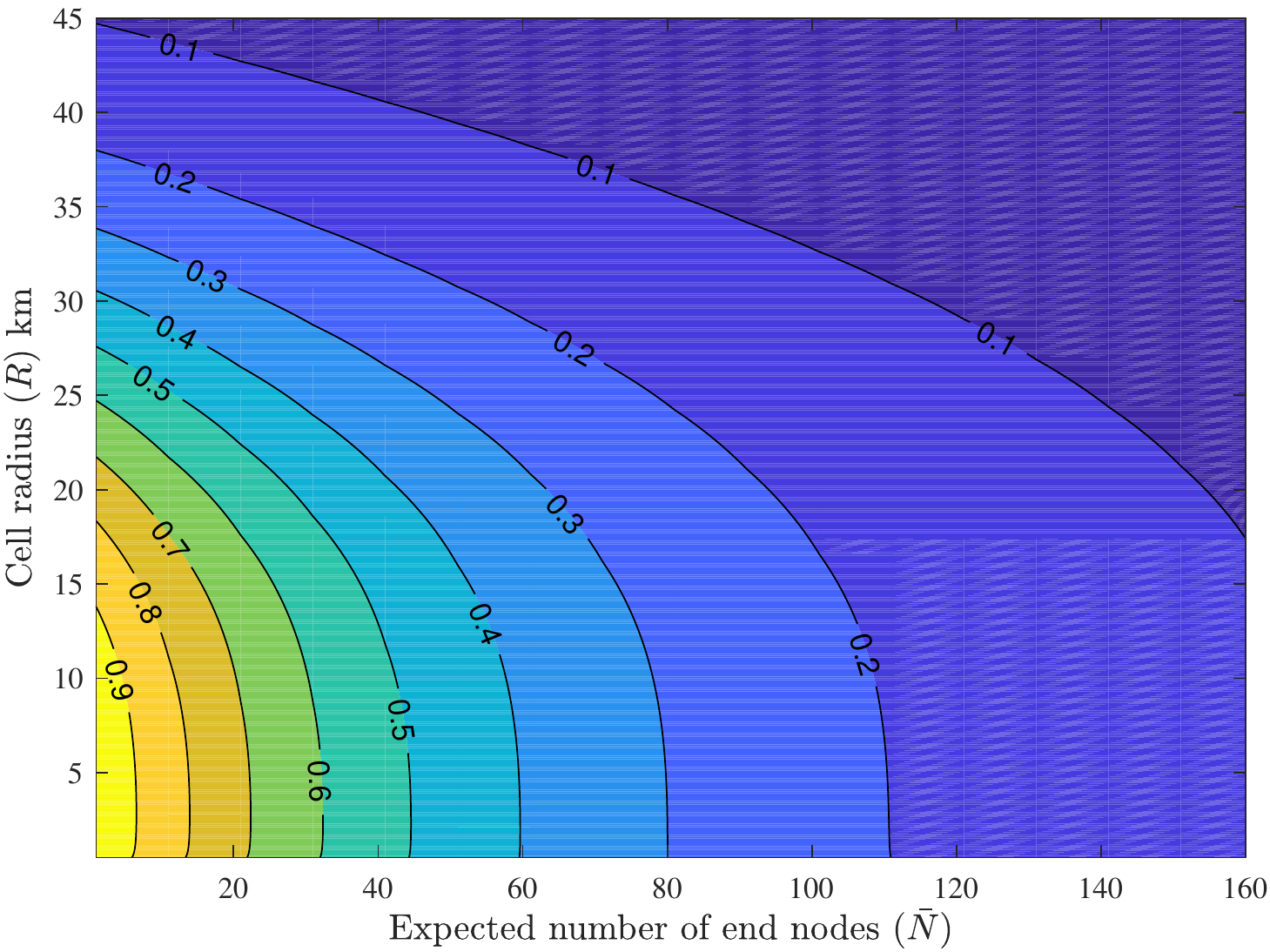}\label{fig:3c}}	
				\vspace{-4pt}
  \caption{Contours of joint coverage probability $P_c\!\left[P_{\textrm{SNR}}\!\cdot\! P_\textrm{SIR}^{\Pi}\right]$ for different duty cycle and transmit power configurations at $\eta = 3$: (a) h1.4: EU 868.0--868.6 MHz with three 125 kHz wide channels and the band has 1\% duty cycle limitation, (b) h1.5: EU 868.7--869.2 MHz offers two channels of 125 kHz bandwidth each, and duty cycle limitation on the band is 0.1\%, (c) h1.6: EU 869.4--869.65 MHz has one 125 kHz channel, and duty cycle cannot exceed 10\%.}
	\vspace{-13pt}
  \label{fig:Cont_all} 
\end{figure*}

We also evaluate the network coverage probability, $P_c[\mathcal{Y}]$, under 
the studied success probabilities i.e., $\mathcal{Y} = \{P_\textrm{SNR}, P_
\textrm{SIR}^*,  P_\textrm{SIR}, P_\textrm{SIR}^{\Pi}\}$ using 
\eqref{eq:eq:CovProbMultiAnnuli}, which is shown in Fig.~\ref{fig:CovProb}. 
These results depict the scalability of a LoRa network with an increasing 
number of EDs for two cell size; $R =6$ km in Fig.~\ref{fig:CovProb}(a) and $R = 12$ km 
in Fig.~\ref{fig:CovProb}(b). The coverage results can give important guidelines for network 
dimensioning.

Fig.~\ref{fig:CovProb}, any subplot, shows that the SNR-
based coverage probability 
($P_c[P_\textrm{SNR}]$) is constant as it is independent of the number of 
EDs. In essence, it gives the noise-only coverage characteristics with 
respect to cell size. On the other hand, SIR-based coverage probabilities, 
i.e., 
$P_c[P_\textrm{SIR}^*]$, $P_c[P_\textrm{SIR}]$ and 
$P_c[P_\textrm{SIR}^{\Pi}]$, decrease exponentially with the increase in EDs. 
This diminishing performance is the direct result of the increasing co-SF and 
inter-SF interference, which together make it less likely for the 
desired signal to achieve the desired SIR protection.
From any of these subfigures, it can observed that the coverage probability 
under dominant co-SF interferer only ($P_c[P_\textrm{SIR}^*]$) is optimistic 
compared to the one given by aggregate co-SF interference 
($P_c[P_\textrm{SIR}]$). $P_c[P_\textrm{SIR}^*]$ gives an upper bound on 
co-SF interference and it reduces its tightness with the increase in the device density.
 In comparison to $P_c[P_\textrm{SIR}]$, the 
coverage probability under both the co-SF and inter-SF ($P_c[P_\textrm{SIR}^{\Pi}]$) 
has a larger decay constant. As a result, $P_c[P_\textrm{SIR}^{\Pi}]$ 
decreases much faster with respect to the number of EDs, and the imperfect 
orthogonality together with same SF interference can have up to 15\% lower 
coverage probability than the same SF case only for 1500 devices per LoRa 
channel. 

The impact of cell size on the coverage probability can be 
observed from Fig.~\ref{fig:CovProb}(a) and Fig.~\ref{fig:CovProb}(b). The 
coverage probability under noise $P_c[P_\textrm{SNR}]$, although independent 
of device density, changes proportionally to the cell size. Whereas, the 
SIR-based coverage probabilities $P_c[P_{\mathcal{Y}}]$ remain invariant to 
the change in the cell size because the relative sum interference remains the 
same for a given $\bar{N}$. How noise and interference together limit the 
scalability of a LoRa network can be observed from the joint coverage 
probability $P_c[P_{\textrm{SNR}}\cdot \mathcal{Y}]$ curves in 
Fig.~\ref{fig:CovProb}(a) and Fig.~\ref{fig:CovProb}(b). From these 
results, it can be concluded that if an EDs achieves full coverage in absence of any interference, 
the coverage probability under both the co-SF and inter-SF interference mainly determines 
the network scalability with an increasing number of EDs.          

%The joint coverage probability $P_c[P_{\textrm{SNR}}\cdot \mathcal{Y}]$
%can also be observed from \textcolor{red}{Fig.~\ref{fig:CovProb}(a) and Fig.~\ref{fig:CovProb}(b)}, for two different cell 
%radii. It is shown how 
%noise and interference together limit the scalability of a LoRa network. 
%Given that an EDs achieves full coverage in absence of any interference, 
%the coverage results under co-SF and inter-SF interference mainly determine 
%the scalability with an increasing number of EDs. 

Fig.~\ref{fig:Cont_all} shows the contour of joint coverage probability $P_c 
\left[P_{\textrm{SNR}}\cdot P_\textrm{SIR}^{\Pi}\right]$ for possible channel 
configurations of a LoRa network with respect to duty cycle and transmit power. These settings 
correspond to h1.4, h1.5 and h1.6 subbands of ERC recommendations 
\cite{ERCrecom2017}, and can serve as a useful indicator for network 
dimensioning. 
For example, if a smart application, such as smart metering, requires a 
certain coverage 
probability which translates directly into quality of service, then for a 
given cell radius, the 
number of EDs operating on a frequency can be determined or suggested. 
Fig.~\ref{fig:Cont_all}(a) and Fig.~\ref{fig:Cont_all}(b) show that as the duty 
cycle 
increases, the number of devices that can achieve a certain coverage 
probability with a given cell size decreases significantly. On the other 
hand, Fig.~\ref{fig:Cont_all}(c) shows that the cell radius increases at 
$p_t = 27$ dBm. However due to the 10\% permitted duty cycle, the impact of 
interference under concurrent transmissions becomes severe, which in turn 
reduces the maximum number of EDs drastically. 

\subsection{SF Allocation Strategies}
\label{subsec:SFAStrategy}
So far, our scalability analysis assumes equal-width annuli for 
SFs' allocation, which we refer to as \textit{equal-interval-based} (EIB) 
scheme. To analyze how an SF allocation scheme influences the network 
performance, we compare EIB with two additional SF allocation strategies, 
namely \textit{equal-area-based} (EAB)~\cite{SFMassiveConn} and \textit{path-loss-based} 
(PLB)~\cite{haxhibeqiri2017lora, MMLoRaWAN} schemes. The EAB scheme uses 
equal-area annuli while PLB defines annuli based on a path loss model and the 
SF-specific SNR thresholds. Using the path loss model, the PLB scheme calculates the SNR with 
respect to the distance. The distance at which 
SNR falls 
below the threshold for the lowest SF defines the outer boundary of the annulus, and 
the higher SF's allocation begins from that boundary and so on. 

\begin{figure}[!t] 
    \centering
  \subfloat[]{%
       \includegraphics[width=0.50\linewidth]{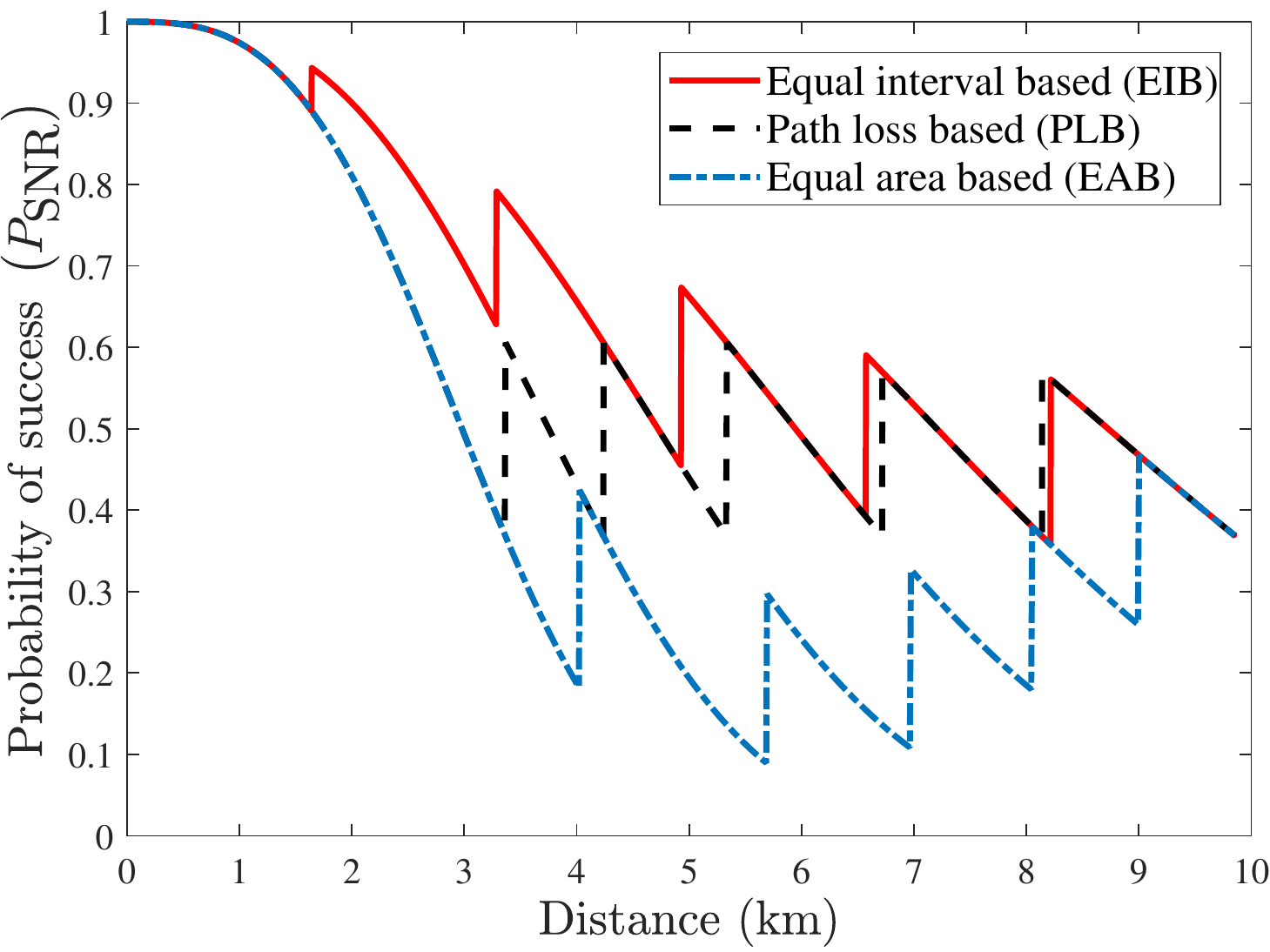}\label{fig:4a}}
  \subfloat[]{%
        \includegraphics[width=0.50\linewidth]{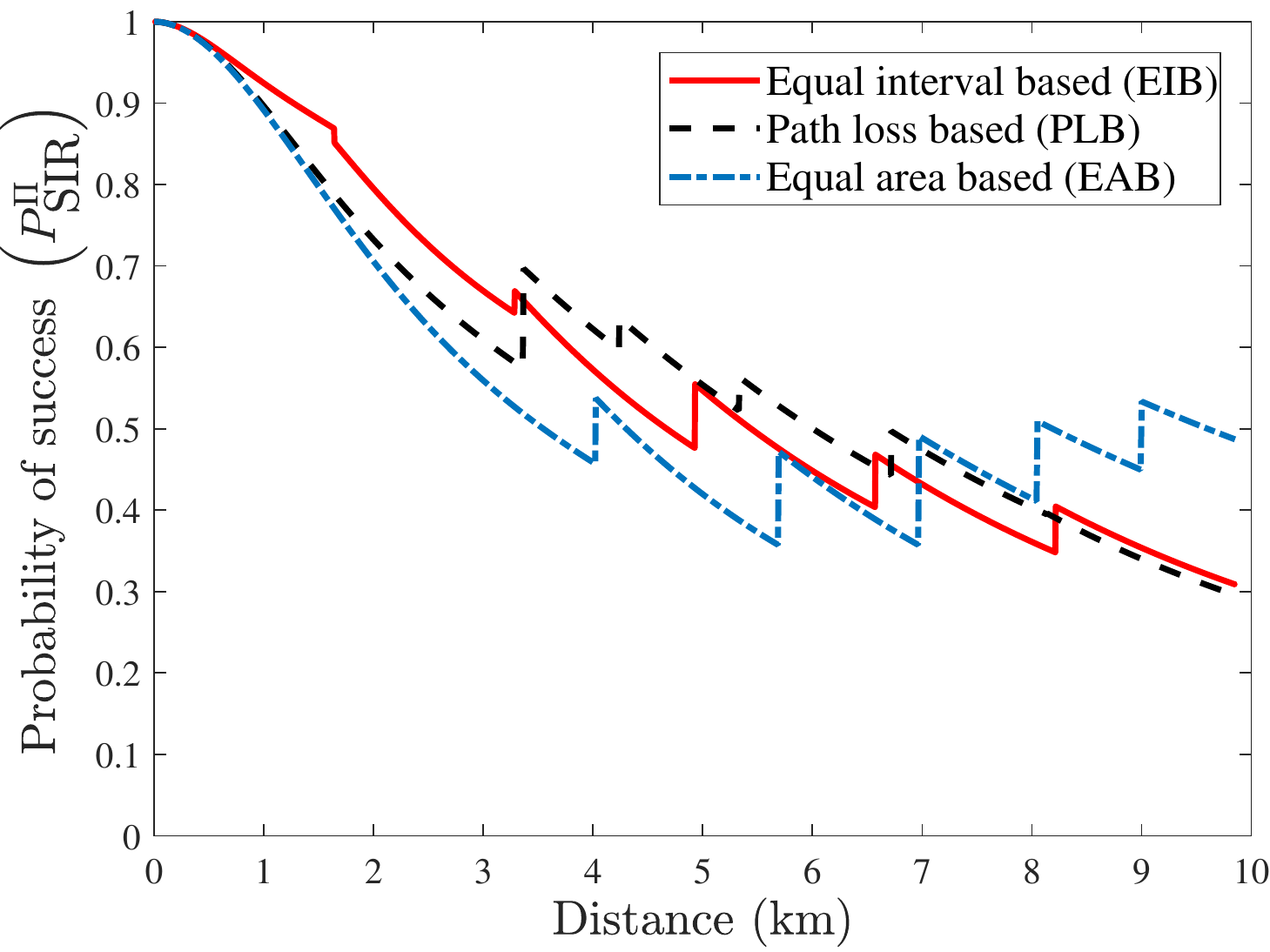}\label{fig:4b}}\hfill
								\vspace{-1pt}
	  \subfloat[]{%
        \includegraphics[width=1\linewidth]{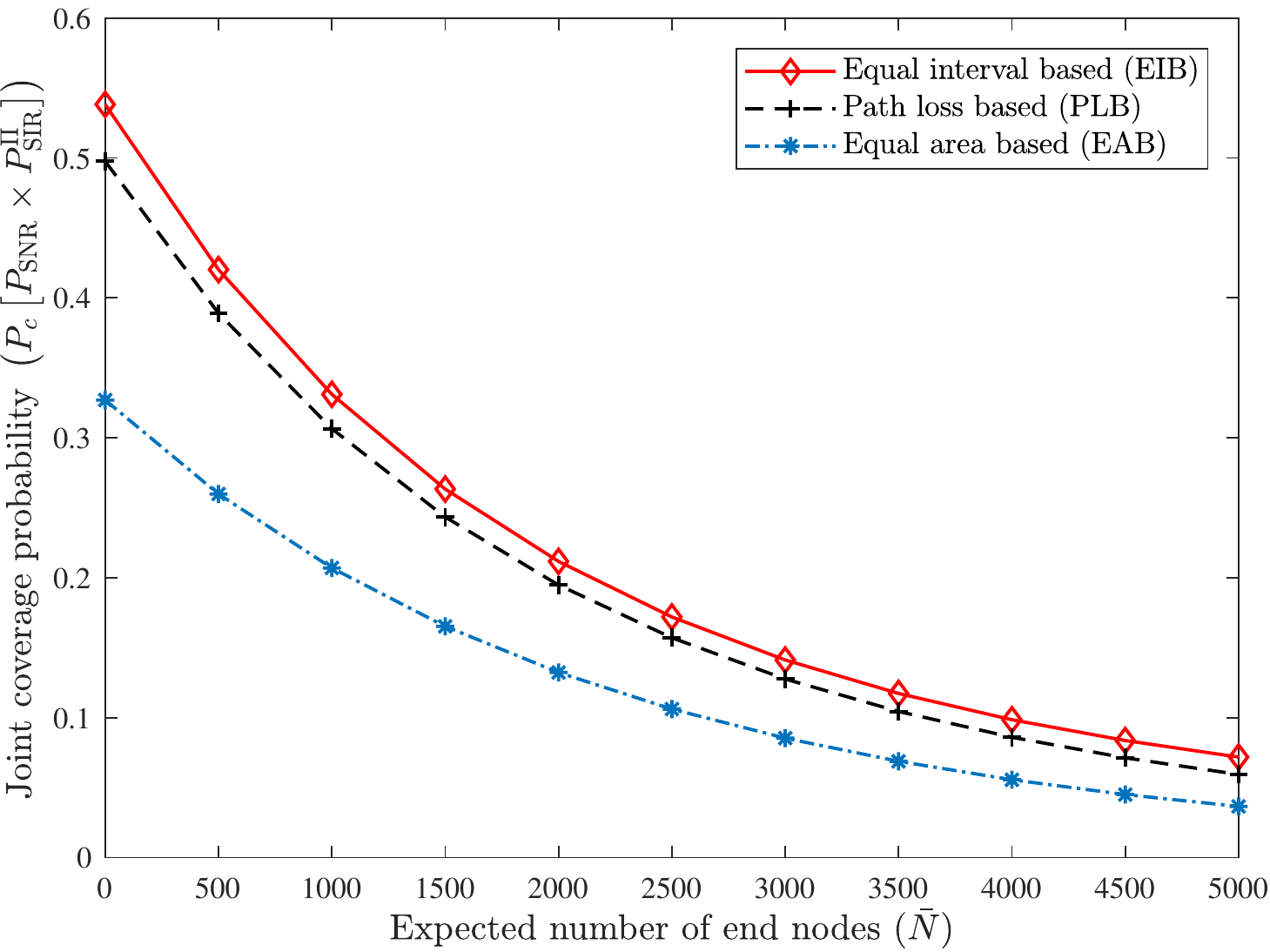}\label{fig:4c}}	
						\vspace{-5pt}
  \caption{Impact of SF allocation schemes on success and coverage probability at $\eta = 3$: (a) success probability under noise, (b) success probability under co-SF and inter-SF interference, (c) Joint coverage probability under noise and interference. Figures (a) and (b) are obtained with $\bar{N}=1500$.}
	\vspace{-14pt}
  \label{fig:SFAllocation}      
\end{figure}

The annuli parameters for each scheme can be determined from 
Table~\ref{tab:SFAllocations}. For PLB scheme, the path loss model is  
defined in Sec.~\ref{subsec:SignChanModel} and the   
SF-specific thresholds $\theta_{\textrm{SF}}$ are given in 
Table~\ref{tab:Tab2}. The success and coverage probabilities obtained under these schemes 
are compared in Fig.~\ref{fig:SFAllocation}. A common cell 
radius $R=9.86$ km, determined by PLB strategy, is used that corresponds to the maximum distance at which the 
required SNR for the highest SF is satisfied.

\bgroup
\def\arraystretch{1.4}
%  1 is the default, change whatever you need
%\setlength{\tabcolsep}{pt}
\begin{table}[!t]
%\medium
\centering
 \caption{Parameters of SF Allocation Schemes}
\label{tab:SFAllocations}
\centering
\begin{tabular}{L{0.56cm}|p{1.91cm}p{1.6cm}l}
\noalign{\hrule height 1pt}
\textbf{Parm.}  & \textbf{EIB} & \textbf{EAB} & \textbf{PLB}\\
\noalign{\hrule height 1pt}
Width   & $r_i \!=\! R/K$  & $r_i \!=\! \ell_i - \ell_{i-1}$ & $r_i \!=\! \ell_i - \ell_{i-1}$\\
        &              & $\ell_i \!=\! R\sqrt{i/6}$      &  $\ell_i \!=\! \{d\!: \textrm{SNR}(d)\!\geq \!\theta_{\textrm{SF}_i}\}$ \\
Area    & $a_i \!=\! \pi r_i^2 (2i-1)$ 	 & ---           & --- \\
\noalign{\hrule height 1pt}
\end{tabular}
\vspace{-10pt}
\end {table}
\egroup

The SNR-based success probability (Fig.~\ref{fig:SFAllocation}
(a)) for EIB 
scheme is mostly higher than that for PLB and EAB schemes except at the cell 
boundary, where the same SF is utilized by each scheme, it becomes equal. 
Both PLB and EAB 
select a higher SF at a distance higher than EIB scheme, and this lag 
results into higher drop in their success probabilities. On the other hand, 
interesting observations can be made for SIR-based (with co-SF and inter-SF 
interference) success probability (see Fig.~\ref{fig:SFAllocation}(b)): a) compared to EIB, the higher area--implying higher number of co-SF interferers--and higher width--meaning more pronounced near-far conditions--of 
the first PLB and EAB annuli causes more performance loss,
%as the desired device moves towards the annulus boundary, 
b) the success 
probability for PLB improves for the subsequent annuli in comparison with 
EIB due to small-area annuli, 
c) the effect of EAB scheme is unusual; it causes performance drop up to a 
certain distance and then performance improves, mainly due to geometrical 
structure of EAB annuli. As the width of an outer annulus is less than an 
inner annulus, the near-far condition in each outer annulus is pronounced 
less. Therefore, after a certain annuli, the probability to achieve co-SF 
SIR-target at the gateway increases as compared to the previous annulus.

Fig.~\ref{fig:SFAllocation}(c) reflects the mentioned drop in success 
probability under PLB and EAB schemes at low SF regions on the joint coverage 
probability. The coverage probability for EIB scheme remains higher than the 
other two.
 %for the studied configuration. 
We also observed (not shown here) that by adding a 
fading margin in PLB scheme, essentially by reducing the cell size, only brings 
the coverage results closer for the studied SF allocation schemes.        

\subsection{Modeling a Multi-Cell LoRa Network}  
In this study, interference modeling of an elemental single-cell LoRa system 
revealed useful scalability results. 
In particular, the coverage  
contours show how to dimension a cell with respect to its size and the number 
of devices, while the numbers are not that optimistic especially if the 
required QoS is high. 
However, in practical applications, the coverage demand is 
expected to span over a large geographical area. As a result, a LoRa network 
will consist of multiple cells to satisfy the coverage and QoS requisites.  
%However, in practical applications, the coverage demand is 
%expected to span over a large geographical area that needs to be served by a network of 
%multiple cells. 
In this respect, interference modeling of a multi-cell network is essential 
which we discuss below based on our proposed approach. 

In smart city applications, the devices are usually clustered with centers at 
the parent points i.e., the gateways. Therefore, a clustering process must be 
defined for interference modeling of a multi-cell network. A 
well-known spatial point process to model the distribution of the gateways is 
Poisson cluster process, where the gateways form a PPP and the devices 
within each cluster form an independent PPP~\cite{MartinInfoTheory}. Next, 
using the joint success probability under noise, intra-cell and inter-cell 
interference, we see how the clustering process affects the 
analysis.

%This type of PCP is called the Mat{\'e}rn cluster process; however, if there 
%is a restriction on the minimum distance between the gateways it is called the 
%Mat{\'e}rn hard-core process. 

Let $\mathcal{L}_{\mathcal{I}_{\textrm{intra}}}^{y_0}(\cdot)$ and $\mathcal{L}
_{
\mathcal{I}_{\textrm{inter}}}(\cdot)$ denote the LT of 
intra-cell interference from $y_0$--the reference cell, and inter-cell 
interference, respectively.
Then, in a multi-cell network, the success probability of a device located at 
$x_1$ in $i$th annulus of $y_0$ is
\begin{align}
P_{\textrm{suc}}(x_{1}, \delta_{ij})  = & \exp\left(-\frac{\sigma\theta_{
\textrm{SF}}}{p_t l(x_1)}\right) \times \prod_{j \in \mathcal{K}} \mathcal{L}_
{\mathcal{I}_{\textrm{intra}}}^{y_0}\left
(\frac{\delta_{ij}}{p_t l(x_{1})}\right) \times \nonumber \\
							 & \prod_{j \in \mathcal{K}} \mathcal{L}_{\mathcal{I}_{\textrm{
inter}}}\left(\frac{\delta_{ij
}}{p_t l(x_{1})}\right)
\label{eq:multicell_suc}
\end{align}
where the first term is $P_{\textrm{SNR}}$ as in \eqref{eq:outage1} while the 
second term is the $P_{\textrm{SIR}}^{\Pi}$ given in \eqref{eq:P_SIR_Total} 
under co-SF interference (i.e., $i=j$) and inter-SF interference in a cell. 
Whereas, the last term considers the impact of interference from 
all the other cells on the success probability.% of the desired transmission. 
At $s = {\delta_{ij}}/{\left(p_t l(x_{1 }\!)\right)}$, $\mathcal{L}_{
\mathcal{I}_{\textrm{inter}}}(s)$ under the PPP distribution of the gateways $
\Phi_G$ can be defined as
\begin{align}
& \mathcal{L}_{\mathcal{I}_{\mathrm{inter}}}(s) = \nonumber \\
& \mathbb{E}_{\Phi_{G},\Phi_{m,j}, G}\!\!\left[\!\exp\!\!\left(\!\!-s 
\!\!\!\!\!\!\!\sum_{y_n\in \Phi_G \!\setminus y_0} \sum_{x_{n,k}\!\in \Phi_{
m,n,j}}\!\!\!\!\!\!\!\!\!{p_t G_{n,k} l\!\left(\!x_{n,k}\!+\!y_n\!\right)}
\!\!\right)\!\!\right]
\label{eq:Laplace_Inter}
\end{align}
where $x_{n,k}+y_n$ is the distance between the reference gateway and the 
device $k$ in annulus $j$ of cell $y_n$, and $G_{n,k}$ is the channel gain of 
the same device.

To evaluate \eqref{eq:Laplace_Inter} where the expectation over $G$ remains 
the same (see \eqref{eq:Laplace_I_CSF11}), one need 
to consider the distance distribution of $x_{n,k}+y_n$ and the clustering 
process of the 'gateways' to find the expectation over $\Phi_{m,j}$ and 
$\Phi_G$ respectively, which we leave as a future work.

%%%%%%%%%%%%%%%%%%%%%%%%%%%%%%%%%%%%%%%%%%%%%%%%%%%%%%%%%%%%%%%%%%%%%%%%%%%
%%%%%%%%%%%%%%%%%%%%%%%%%%%%%%%%%%%%%%%%%%%%%%%%%%%%%%%%%%%%%%%%%%%%%%%%%%%
%%%%%%%%%%%%%%%%%%%%%%%%%%%%%%%%%%%%%%%%%%%%%%%%%%%%%%%%%%%%%%%%%%%%%%%%%%%
%%%%%%%%%%%%%%%%%%%%%%%%%%%%%%%%%%%%%%%%%%%%%%%%%%%%%%%%%%%%%%%%%%%%%%%%%%%
\section{Conclusions}
\label{sec:Conclusions}
%%%%%%%%%%%%%%%%%%%%%%%%%%%%%%%%%%%%%%%%%%%%%%%%%%%%%%%%%%%%%%%%%%%%%%%%%%%
%%%%%%%%%%%%%%%%%%%%%%%%%%%%%%%%%%%%%%%%%%%%%%%%%%%%%%%%%%%%%%%%%%%%%%%%%%%
%%%%%%%%%%%%%%%%%%%%%%%%%%%%%%%%%%%%%%%%%%%%%%%%%%%%%%%%%%%%%%%%%%%%%%%%%%%
%%%%%%%%%%%%%%%%%%%%%%%%%%%%%%%%%%%%%%%%%%%%%%%%%%%%%%%%%%%%%%%%%%%%%%%%%%%
In this paper, we investigated the impact of interference on a LoRa network 
caused by simultaneous transmissions using the same SF as well as different 
SFs. While, the co-SF interference is natural and requires SIR protection to 
have any 
benefits from capture effect, the imperfect orthogonality among SFs can also 
cause a significant impact in high-density deployment of devices. To this 
end, using stochastic geometry to model the interference field, we derived 
the SIR distributions to capture the uplink outage and coverage performance 
with respect to the distance from the gateway. The SIR distributions are 
derived based on the aggregate co-SF and inter-SF interference power. 
The results obtained by comparing the  
aggregate co-SF interference alone to the corresponding 
upper bound based on dominant interferer, defy the validity of the 
argument- \textit{LoRa is impervious to the 
cumulative interference effects}\cite{georgiou2017low}- which is shown here 
to be dependent on the 
device density. 
Moreover, our analysis reveals that the network 
scalability under the joint impact of co-SF and inter-SF interference is 
more accurate compared to the optimistic results usually reported when 
considering co-SF alone. We showed in a LoRa frequency channel only a limited 
number of devices can successfully transmit, otherwise the devices would 
waste energy in retransmissions of collided packets. In particular, for 
higher SFs this effect is more noticeable due to lower success probability.   
We summarized the usefulness of our analytical models for: a) network 
dimensioning under reliability constraints using contour plots for three 
baseline LoRa channel settings, b) interference modeling of a multi-cell LoRa 
network. In addition, we analyzed the network performance for three SF allocation schemes and showed that 
a simple equal-width-based scheme yields better results than the equal-area- 
and path loss-based schemes.
\vspace{-5pt}
%Finally, we presented contour plots for three baseline LoRa channel 
%settings that can be utilized for network dimensioning 
%under reliability constraints.

%%%%%%%%%%%%%%%%%%%%%%%%%%%%%%%%%%%%%%%%%%%%%%%%%%%%%%%%%%%%%%%%%%%%%%%%%
%%%%%%%%%%%%%%%%%%%%%%%%%%%%%%%%%%%%%%%%%%%%%%%%%%%%%%%%%%%%%%%%%%%%%%%%% 

\bibliographystyle{IEEEtran}
\bibliography{bib}

\end{document}